\documentclass{aastex61}
\usepackage{textcomp}

\received{July 1, 2016}
\revised{September 27, 2016}
\accepted{\today}
\submitjournal{ApJ}

\shorttitle{Forbush decrease PAMELA}
\shortauthors{R. Munini et al.}

\usepackage{textcomp}

\begin{document}

\title{Evidence of energy and charge sign dependence of the recovery time for the December 2006 Forbush event measured by the PAMELA experiment}

\correspondingauthor{Riccardo Munini}
\email{riccardo.munini@ts.infn.it}

\author{R. Munini}
\affil{INFN, Sezione di Trieste I-34149 Trieste, Italy}

\author{M.~Boezio}
\affiliation{INFN, Sezione di Trieste I-34149 Trieste, Italy}

\author{A.~Bruno}
\affiliation{INFN, Sezione di Bari, I-70126 Bari, Italy}

\author{E. C. Christian}
\affiliation{Heliophysics Division, NASA Goddard Space Flight Ctr, Greenbelt, MD, USA.}

\author{G. A. de Nolfo}
\affiliation{Heliophysics Division, NASA Goddard Space Flight Ctr, Greenbelt, MD, USA.}

\author{V.~Di~Felice}
\affiliation{INFN, Sezione di Rome ``Tor Vergata'', I-00133 Rome, Italy}
\affiliation{Agenzia Spaziale Italiana (ASI) Science Data Center, I-00044 Frascati, Italy}

\author{M.~Martucci}
\affiliation{University of Rome ``Tor Vergata'', Department of Physics,  I-00133 Rome, Italy}
\affiliation{INFN, Laboratori Nazionali di Frascati, Via Enrico Fermi 40, I-00044 Frascati, Italy}

\author{M. Merge'}
\affiliation{INFN, Sezione di Rome ``Tor Vergata'', I-00133 Rome, Italy}
\affiliation{University of Rome ``Tor Vergata'', Department of Physics,  I-00133 Rome, Italy}

\author{I. G. Richardson}
\affiliation{Heliophysics Division, NASA Goddard Space Flight Ctr, Greenbelt, MD, USA.}
\affiliation{GPHI and Department of Astronomy, University of Maryland, College Park.}

\author{J. M. Ryan}
\affiliation{Space Science Center, University of New Hampshire, Durham, NH, USA.}

\author{S. Stochaj }
\affiliation{Electrical and Computer Engineering, New Mexico State University, Las Cruces, NM, USA.}

\author{O.~Adriani}
\affiliation{University of Florence, Department of Physics, I-50019 Sesto Fiorentino, Florence, Italy}
\affiliation{INFN, Sezione di Florence, I-50019 Sesto Fiorentino, Florence, Italy}

\author{G.~C.~Barbarino}
\affiliation{University of Naples ``Federico II'', Department of Physics, I-80126 Naples, Italy}
\affiliation{INFN, Sezione di Naples,  I-80126 Naples, Italy}

\author{G.~A.~Bazilevskaya}
\affiliation{Lebedev Physical Institute, RU-119991, Moscow, Russia}

\author{R.~Bellotti}
\affiliation{University of Bari, Department of Physics, I-70126 Bari, Italy}
\affiliation{INFN, Sezione di Bari, I-70126 Bari, Italy}

\author{M.~Bongi}
\affiliation{University of Florence, Department of Physics, I-50019 Sesto Fiorentino, Florence, Italy}
\affiliation{INFN, Sezione di Florence, I-50019 Sesto Fiorentino, Florence, Italy}

\author{V.~Bonvicini}
\affiliation{INFN, Sezione di Trieste I-34149 Trieste, Italy}

\author{S.~Bottai}
\affiliation{INFN, Sezione di Florence, I-50019 Sesto Fiorentino, Florence, Italy}

\author{F.~Cafagna}
\affiliation{University of Bari, Department of Physics, I-70126 Bari, Italy}
\affiliation{INFN, Sezione di Bari, I-70126 Bari, Italy}

\author{D.~Campana}
\affiliation{INFN, Sezione di Naples,  I-80126 Naples, Italy}

\author{P.~Carlson}
\affiliation{KTH, Department of Physics, Oskar Klein Centre for Cosmoparticle Physics, AlbaNova University Centre, SE-10691 Stockholm, Sweden}

\author{M.~Casolino}
\affiliation{INFN, Sezione di Rome ``Tor Vergata'', I-00133 Rome, Italy}
\affiliation{RIKEN, Advanced Science Institute, Wako-shi, Saitama, Japan}

\author{G.~Castellini}
\affiliation{IFAC, I-50019 Sesto Fiorentino, Florence, Italy}

\author{C.~De~Santis}
\affiliation{INFN, Sezione di Rome ``Tor Vergata'', I-00133 Rome, Italy}

\author{A.~M.~Galper}
\affiliation{National Research Nuclear University MEPhI, RU-115409 Moscow}

\author{A.~V.~Karelin}
\affiliation{National Research Nuclear University MEPhI, RU-115409 Moscow}

\author{S.~V.~Koldashov}
\affiliation{National Research Nuclear University MEPhI, RU-115409 Moscow}

\author{S.~Koldobskiy}
\affiliation{National Research Nuclear University MEPhI, RU-115409 Moscow}

\author{S.~Y.~Krutkov}
\affiliation{Ioffe Physical Technical Institute,  RU-194021 St. Petersburg, Russia}

\author{A.~N.~Kvashnin}
\affiliation{Lebedev Physical Institute, RU-119991, Moscow, Russia}

\author{A.~Leonov}
\affiliation{National Research Nuclear University MEPhI, RU-115409 Moscow}

\author{V.~Malakhov}
\affiliation{National Research Nuclear University MEPhI, RU-115409 Moscow}

\author{L.~Marcelli}
\affiliation{INFN, Sezione di Rome ``Tor Vergata'', I-00133 Rome, Italy}

\author{A.~G.~Mayorov}
\affiliation{National Research Nuclear University MEPhI, RU-115409 Moscow}

\author{W.~Menn}
\affiliation{Universit\"{a}t Siegen, Department of Physics, D-57068 Siegen, Germany}

\author{V.~V.~Mikhailov}
\affiliation{National Research Nuclear University MEPhI, RU-115409 Moscow}

\author{E.~Mocchiutti}
\affil{INFN, Sezione di Trieste I-34149 Trieste, Italy}

\author{A.~Monaco}
\affiliation{University of Bari, Department of Physics, I-70126 Bari, Italy}
\affiliation{INFN, Sezione di Bari, I-70126 Bari, Italy}

\author{N.~Mori}
\affiliation{INFN, Sezione di Florence, I-50019 Sesto Fiorentino, Florence, Italy}

\author{G.~Osteria}
\affiliation{INFN, Sezione di Naples,  I-80126 Naples, Italy}

\author{B.~Panico}
\affiliation{INFN, Sezione di Naples,  I-80126 Naples, Italy}

\author{P.~Papini}
\affiliation{INFN, Sezione di Florence, I-50019 Sesto Fiorentino, Florence, Italy}

\author{M.~Pearce}
\affiliation{KTH, Department of Physics, Oskar Klein Centre for Cosmoparticle Physics, AlbaNova University Centre, SE-10691 Stockholm, Sweden}

\author{P.~Picozza}
\affiliation{INFN, Sezione di Rome ``Tor Vergata'', I-00133 Rome, Italy}
\affiliation{University of Rome ``Tor Vergata'', Department of Physics,  I-00133 Rome, Italy}

\author{M.~Ricci}
\affiliation{INFN, Laboratori Nazionali di Frascati, Via Enrico Fermi 40, I-00044 Frascati, Italy}

\author{S.~B.~Ricciarini}
\affiliation{IFAC, I-50019 Sesto Fiorentino, Florence, Italy}

\author{M.~Simon}
\affiliation{Universit\"{a}t Siegen, Department of Physics, D-57068 Siegen, Germany}

\author{R.~Sparvoli}
\affiliation{INFN, Sezione di Rome ``Tor Vergata'', I-00133 Rome, Italy}
\affiliation{University of Rome ``Tor Vergata'', Department of Physics,  I-00133 Rome, Italy}

\author{P.~Spillantini}
\affiliation{University of Florence, Department of Physics, I-50019 Sesto Fiorentino, Florence, Italy}
\affiliation{INFN, Sezione di Florence, I-50019 Sesto Fiorentino, Florence, Italy}

\author{Y.~I.~Stozhkov}
\affiliation{Lebedev Physical Institute, RU-119991, Moscow, Russia}

\author{A.~Vacchi}
\affiliation{INFN, Sezione di Trieste I-34149 Trieste, Italy}
\affiliation{University of Udine, Department of Mathematics and Informatics, I-33100 Udine, Italy}

\author{E.~Vannuccini}
\affiliation{INFN, Sezione di Florence, I-50019 Sesto Fiorentino, Florence, Ital}

\author{G.~Vasilyev}
\affiliation{Ioffe Physical Technical Institute,  RU-194021 St. Petersburg, Russia}

\author{S.~A.~Voronov}
\affiliation{National Research Nuclear University MEPhI, RU-115409 Moscow}

\author{Y.~T.~Yurkin}
\affiliation{National Research Nuclear University MEPhI, RU-115409 Moscow}

\author{G.~Zampa}
\affiliation{INFN, Sezione di Trieste I-34149 Trieste, Italy}

\author{N.~Zampa}
\affiliation{INFN, Sezione di Trieste I-34149 Trieste, Italy}

\author{M. S. ~Potgieter}
\affiliation{North-West University, Centre for Space Research,2520 Potchefstroom, South Africa}



\begin{abstract}

New results on the short-term galactic cosmic ray (GCR) intensity variation (Forbish decrease) 
in December $2006$   
measured by the PAMELA instrument are presented. 
Forbush decreases are sudden suppressions of the GCR intensities which are
associated with the passage of interplanetary transients such as shocks and interplanetary coronal mass ejections (ICMEs). Most of the past 
measurements of this phenomenon were carried out with ground-based detectors such as neutron monitors or muon 
telescopes. These techniques allow only the indirect detection of the overall GCR intensity over an 
integrated energy range. For the first time, thanks to the unique features of the 
PAMELA magnetic spectrometer, the Forbush decrease commencing on $2006$ December 14, following a CME at the Sun on $2006$
December  $13$  was studied in a wide rigidity range
($0.4-20$ GV) and for different species of GCRs detected directly in space. 
The daily averaged GCR proton intensity was used to investigate the rigidity dependence 
of the amplitude and the recovery time of the Forbush decrease.
Additionally, for the first time, the temporal variations in the helium and electron intensities during a Forbush decrease were studied. 
Interestingly, the temporal evolutions of the helium and
proton intensities during the Forbush decrease were found in good agreement, while the low rigidity electrons
($< 2$ GV) displayed a faster recovery. This difference in the electron recovery is interpreted as a charge-sign dependence introduced by drift
motions experienced by the GCRs during their propagation through the heliosphere.

\end{abstract}



\section{Introduction} \label{sec:intro}

The solar environment significantly affects the spectrum of galactic cosmic rays (GCRs) observed at Earth below a few tens of GV. 
Before reaching the Earth, GCRs propagate through the heliosphere, the 
region of space formed by the continuous outflow of plasma from the solar corona,
also known as the solar wind (SW). In addition, the magnetic field of the Sun	
 freezes into the solar wind plasma and is transported through the heliosphere, forming the so called 
 heliospheric magnetic field (HMF) \citep{HMFcite}. 
The GCRs, traveling through the interplanetary medium, interact
with the SW and the HMF. As a consequence their spectra are modified in intensity and shape with respect 
to the local interstellar spectrum (LIS) \citep{pot13a}. 
In addition, in response to the $11$-years solar cycle \citep{solcycle}, a long-term modulation of the GCRs is observed. 
The solar modulation of GCR is anti-correlated with respect to the solar cycle since the 
particle fluxes reach their maximal intensity during periods of low solar activity.    

On top of the long-term solar modulation, short-term modulation effects also occur.
For example, the GCR intensity may be modulated by transient phenomena 
as interplanetary coronal mass ejections (e.g. \cite{CME1,CME2} and references therein). These ICMEs consist of  magnetized coronal plasma 
ejected from the Sun's surface that then propagates through the heliosphere. Some ICMEs propagate through the solar wind 
at super-Alfvenic speeds and drive a shock ahead of them  (e.g. \cite{Jian}). 
As the ICME passes near Earth the sheath following the shock acts as a shield against the ambient 
population of GCRs since they cannot easily diffuse through the region of enhanced turbulence in the sheath
\citep{wiz}. Moreover, as it propagates from the Sun to the Earth, the ICME itself
is progressively populated by GCRs that perpendicularly diffuse into the magnetic cloud (e.g. \cite{richcane,arunbabu,cloud1,cloud2}). 
As an overall effect a sudden suppression of GCR intensity is observed. 
Such a phenomenon initially identified by \cite{forbush} (and also by  \cite{Hess}) 
and hence called a Forbush decrease, can last up to several days suppressing the GCR intensity up to  
about $30\% - 40 \%$ or even more (e.g. \cite{Cane2000} and references therein).
The relative contributions of shocks and ICMEs in causing Forbush decreases is still a matter of debate, 
and likely varies from event to event and observationally depends on the trajectory of the observer through the shock 
and ICME (e.g., Figure 1 of \cite{CME2}).
In addition, recurrent short-term GCR decreases have been measured
in association with the passage of corotating interaction regions (CIRs). Such regions of compressed plasma, formed
at the leading edges of high-speed solar wind streams originating from
coronal holes and interacting with the preceding slow solar wind, 
are a well known cause of periodic CR decreases (e.g. \cite{simpson,Richardson} and references therein). 
The study of the CIR associated GCR intensity decreases with the PAMELA data 
will be the subject of a future paper.



\begin{figure}[]
 \centering
 \includegraphics[width=19.5cm]{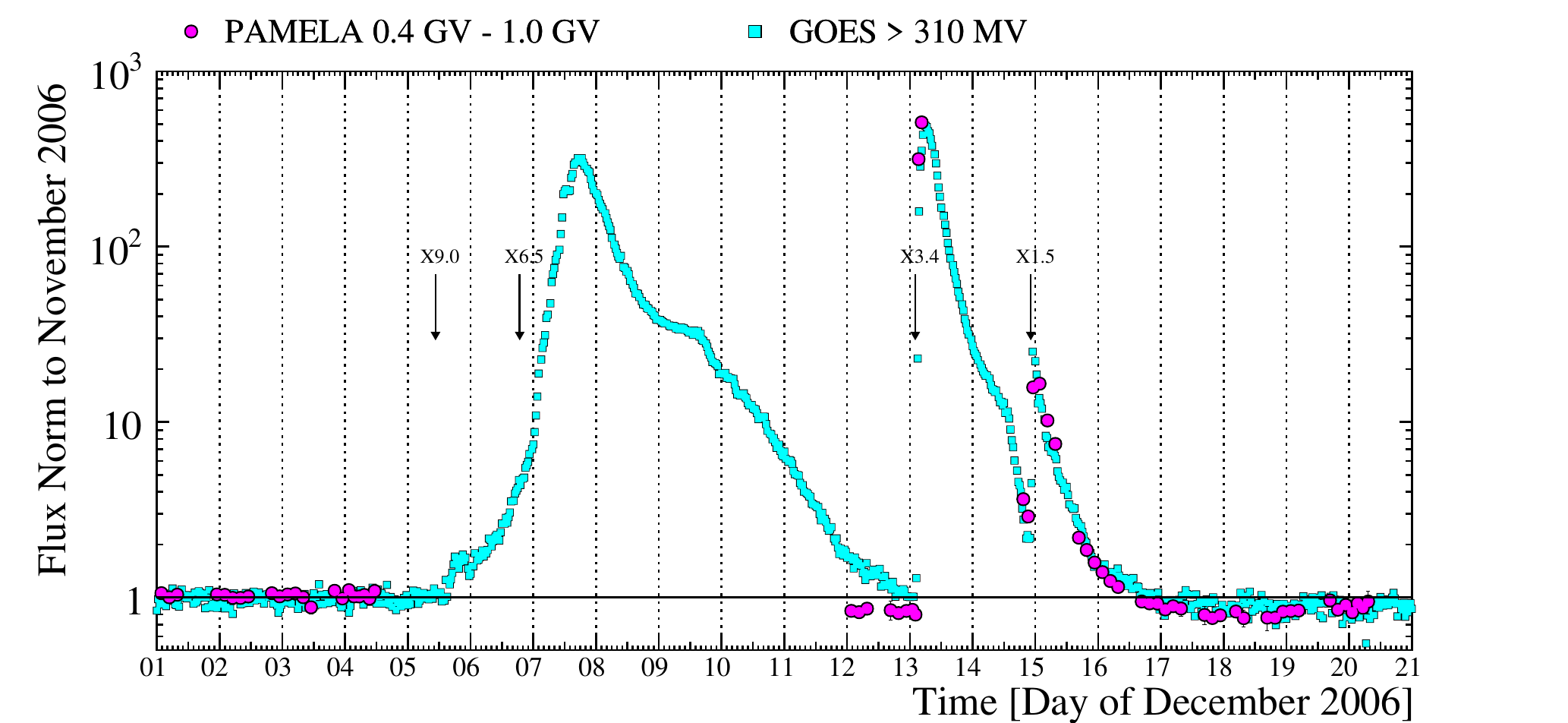}
 \caption{The PAMELA proton intensity (full circles) measured between the $2006$ December 1 and 21
 in the rigidity range $0.4$ - $1.0$ GV.  The horizontal line represents the GCR 
 reference intensity to which the data were normalized, i.e. the average intensity of GCRs proton measured during  November $2006$. Each point represents
 three hours of data taking. Missing data are due to maintenance procedure and on-board system reset of the satellite. 
 The integrated GOES-12 proton data $>310$ MV (full squares) averaged over $30$ minutes are also showed for comparison. Data were taken from 
 \url{http://satdat.ngdc.noaa.gov/sem/goes/data/new_full} and were normalized to the average  November $2006$  GOES-12 proton intensity. 
 PAMELA and GOES-12 data show the $2006$ December 13 and 14 SEP events registered as a sudden increase of the proton 
 intensity. GOES-12 data also shows the $2006$ December 5 and 6 event. The flare time at the Sun is indicating by the vertical arrows.}
 \label{fig1}
\end{figure}

The Forbush decrease, observed by the PAMELA space mission discussed in this work, occurred in December 2006, during the 
extraordinary deep and prolonged solar minimum between solar cycles 23 and 24 \citep{potgieter_prot,Russel}.
Solar minimum periods are particularly interesting to 
measured transient phenomena like Forbush decrease. Being the Sun's activity at minimum, the overall structure of the HMF is well ordered 
and easier to reproduce from a modeling point of view. Moreover the time variation of GCR intensity is slower with
respect to a period of high Sun's activity. Solar minima are thus well suited to study and disentangle the relative contribution 
to the GCR intensity variations due to ICME and shocks from that due to solar modulation.  
Remarkably, the minimum between solar cycles 23 and 24 was characterized by very stable heliospheric conditions,
except for the powerful solar events that occurred during December $2006$.
Four X class solar flares originated during  December $2006$ as solar active region $10930$
rotated across the visible hemisphere of the Sun. 
The first of these X-class flares (X$9.0$) occurred on $2006$  
December $5$ at  E$79$\textdegree \ with peak emission  at  
$1035$  UT and was followed by an X$6.5$ flare on the $2006$ December $6$ at E$63$\textdegree, with peak emission at $1847$ UT. 
On the $2006$ December $13$ another X$3.4$ flare occurred at W$23$\textdegree \ with peak emission at  $0240$ UT followed by 
a X$1.5$ flare on the  $2006$ December $14$ at W$46$\textdegree \  with peak emission at $2215$ UT \citep{solfl2006}.
These events produced an enhancement  
of particles up to several GV that was recorded and extensively studied by the PAMELA 
  instrument \citep{pam_2006_flare} and other satellites.
  The PAMELA instrument also measured the variations of the geomagnetic cutoff latitude as a function of rigidity during the $2006$ December $14$ magnetospheric
  storm caused by the ICME associated with the December $13$ event \citep{cutoffvar}. Figure \ref{fig1} shows the 
proton intensity ($400$-$1000$ MV) measured in  December $2006$ by the PAMELA instrument (full circles). 
Data are normalized to the   November $2006$ average proton intensity that was considered as the GCR background level. 
Each point represents three hours of 
data taking. For comparison the GOES-12 proton integrated data ($>310$ MV) are shown 
(full squares). The GOES-12 data were also normalized to the   November $2006$ average proton intensity. 
The PAMELA data exhibit a sudden increase in the proton intensity associated with the X3.4 flare on 13 December.
However, due to a scheduled 
maintenance procedure, no data were collected during the 5/6 December events.
The GOES-12 data show an increase of the proton intensity corresponding 
to all four of the X-class flares in December $2006$. In addition, halo CMEs were observed by the LASCO
coronagraphs on SOHO in association with the events of $2006$ December 13 and 14, with speeds of $1774$ km/s 
and $1042$ km/s $($data taken from \url{https://cdaw.gsfc.nasa.gov/CME_list/
}$)$ 
respectively while the $2006$ December 5 and 6 events occurred during SOHO/LASCO data gaps. 

The passage of the $2006$ December $13$ CME caused a Forbush decrease that lasted for several days
which is evident in Figure 1 and will be shown in greater detail below.
Thanks to its quasi-polar orbit the PAMELA instrument has measured this event in the rigidity
range from $400$ MV to $20$ GV.  This extends and completes
studies based on other measurements,  typically performed on the ground,  either
by neutron monitors or muon telescopes (e.g. \cite{muon,neutron}). The performance of these ground-based detectors is 
limited since they can only determine an integral flux above an energy threshold that depends 
on the latitudinal geomagnetic cutoff at the location of the monitor. For the first time a Forbush decrease was extensively studied with 
GCRs detected directly in space in a wide rigidity range. 
The accuracy of the rigidity reconstruction and the high counting statistics allowed the rigidity-dependences of 
the amplitude and the recovery time of this event to be studied.
In addition, the PAMELA instrument allowed the temporal evolution of the GCRs to be studied for 
several particle species. In particular the galactic cosmic ray proton, helium and 
electron intensities over time were studied. 
By comparing GCRs with oppositely signed charges, it is possible to identify differences in the Forbush decrease 
amplitude and recovery time that could be introduced by drift motions 
experienced by the GCRs during their interaction with and propagation through the ICME \citep{luo}.
After a brief discussion of the PAMELA instrument in Section \ref{sec:isntr} the data analysis will be 
discussed in Section \ref{sec:data} and the results will be presented in Section \ref{sec:res}.

\section{The PAMELA instrument}
\label{sec:isntr}

PAMELA, 
Payload for Antimatter-Matter Exploration and Light-nuclei Astrophysics, is a satellite-borne experiment
designed to make long duration measurements of the cosmic
radiation from Earth orbit \citep{pic07}. 
The instrument collected GCRs in space for almost ten years from the $2006$ June $15$  
when it was launched from the Baikonur cosmodrome in Kazakhstan, to late  January $2016$. 
Until  September $2010$ the orbit was elliptical  with  altitudes ranging
between $350$ and $610$ km with an inclination of $70$ degrees. After $2010$ the 
orbit was changed and became circular at a constant altitude of $570$ km. 

\begin{figure}[t!]
 \centering
 \includegraphics[scale=0.82]{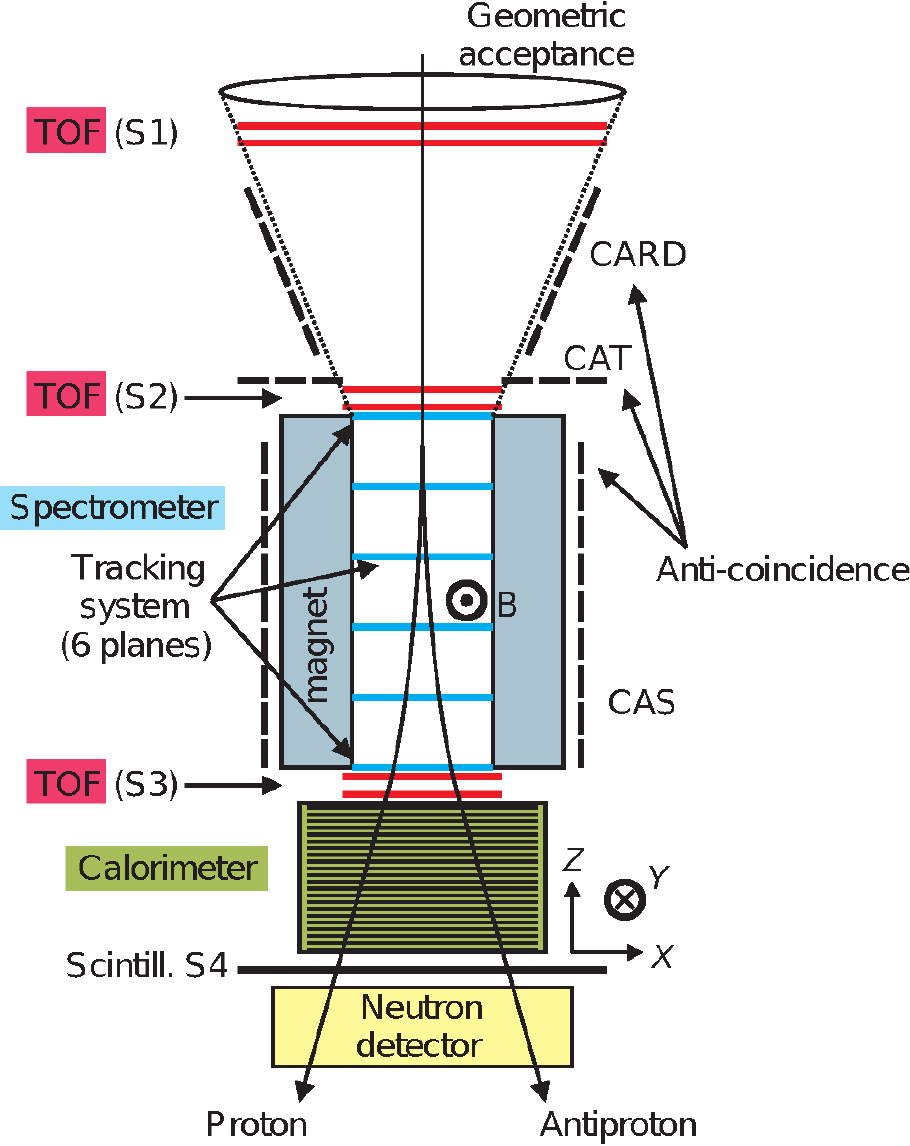}
 \caption{PAMELA and its sub-detectors.}
 \label{fig2}
\end{figure}

The apparatus is schematically shown in Figure \ref{fig2}.
The core of the instrument is the magnetic spectrometer \citep{trk}, 
a silicon tracking system in the $0.43$~T magnetic field generated by a permanent magnet. 
The $300 \; \mu$m thick double-sided Si sensors of
the tracking system measure two independent impact coordinates (bending X-view and non-bending Y-view) on
each plane, accurately reconstructing the particle deflection, measuring its rigidity (momentum divided by
charge) with a maximum detectable rigidity of
$1.2$~TV, and the sign of the electric
charge. The instrument geometric factor, as defined by the 
magnetic cavity, is $21.5$ cm$^2$ sr. A system of six layers of plastic scintillators, arranged in three double planes (S1, S2 and S3), provides
 a fast signal for triggering the data acquisition. Moreover it
contributes to particle identification measuring 
 the ionization energy loss and the Time
of Flight (ToF) of traversing particles with a resolution of
$300$ ps, assuring charge particle absolute value determination and albedo particle\footnote{Particles produced in cosmic-ray
interactions with the atmosphere with rigidities lower
than the geomagnetic cutoff that, propagating along Earth's magnetic field line, re-enter the atmosphere
in the opposite hemisphere but at a similar magnetic latitude.} rejection \citep{ToF}. 
The hadron-lepton discrimination is provided by an electromagnetic imaging W/Si calorimeter,
$16.3$ radiation lengths and $0.6$ interaction lengths deep \citep{calo}.
Thanks to its longitudinal and transverse segmentation, the
calorimeter exploits the different development of electromagnetic and hadronic showers, allowing a rejection power
of interacting and non-interacting hadrons at the order of
$10^5$. A neutron counter \citep{ND}
contributes to discrimination power by detecting the increased neutron production
 in the calorimeter associated with hadronic showers compared to electromagnetic ones,
 while a plastic scintillator, placed beneath the calorimeter,
increases the identification of high-energy electrons. 
The whole instrument is surrounded by an anticoindence
system (AC) of three scintillators (CARD, CAS and CAT)
for the rejection of background events \citep{AC}. For a complete review of the PAMELA apparatus see \citep{PAMphysrep,nuovocimento}.

\section{Data analysis}
\label{sec:data}

\begin{figure}[t!]
\begin{center}

\includegraphics [height=59mm] {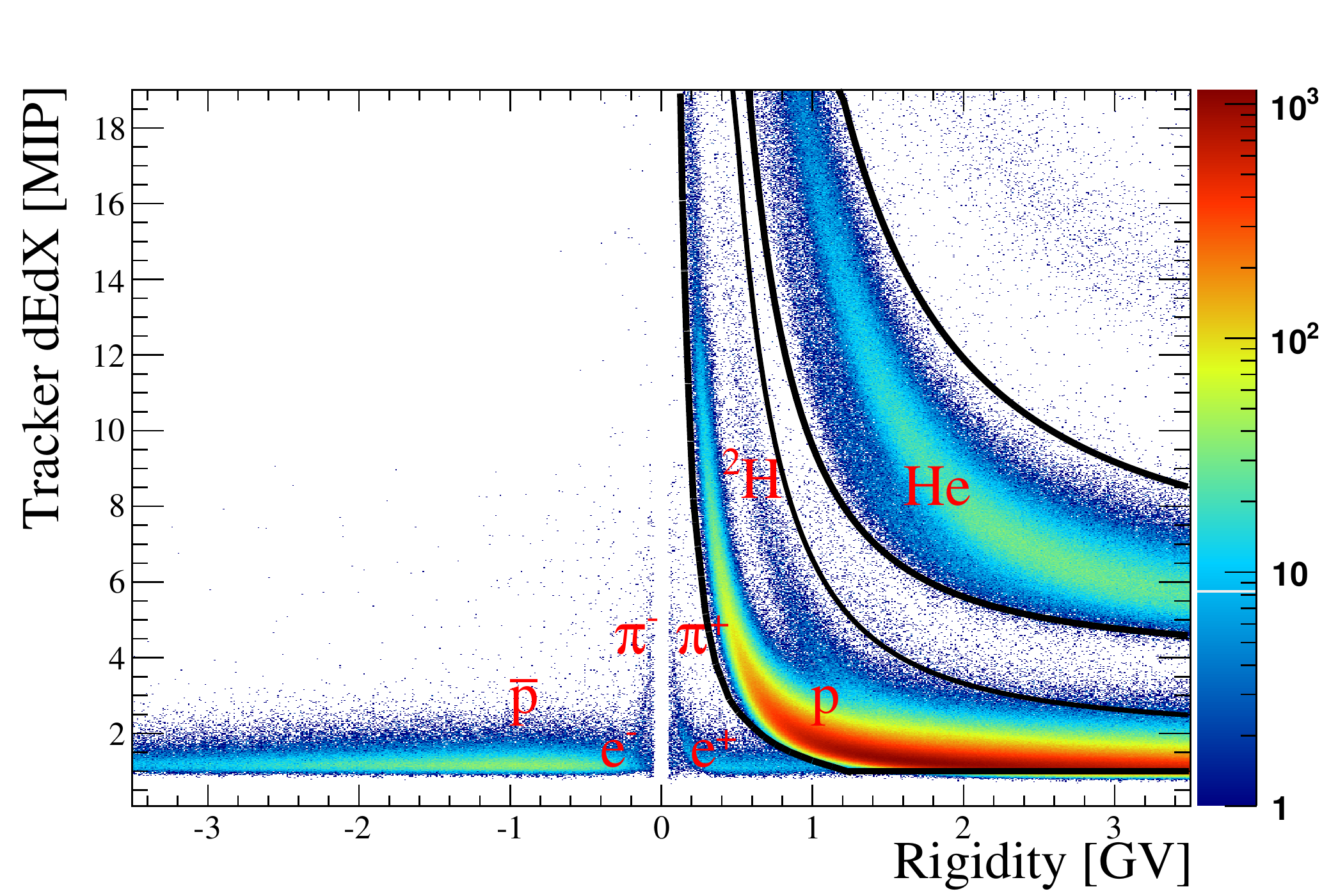}
\includegraphics [height=59mm] {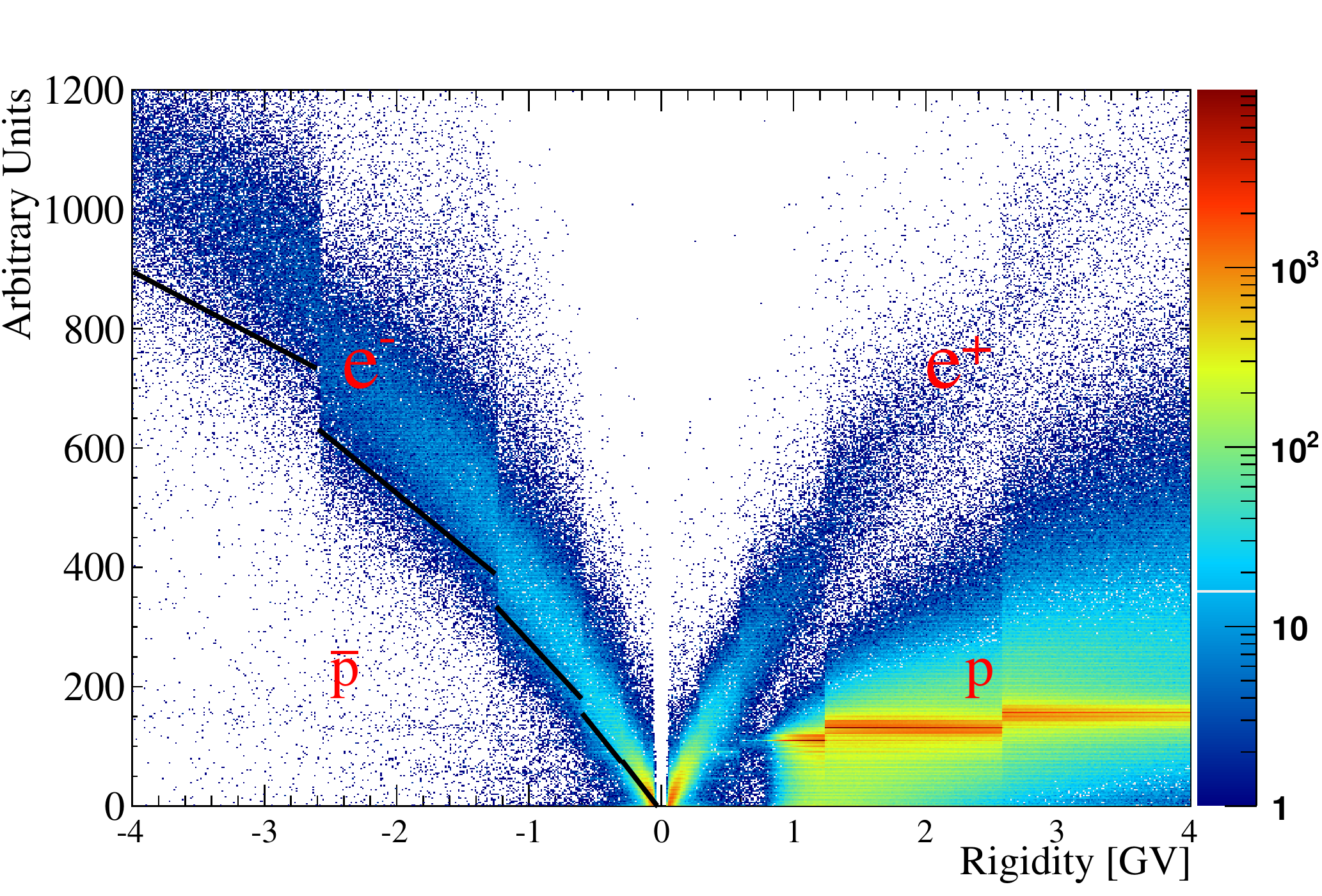}

\end{center}
\caption{Left panel: the average ionization energy losses on the tracker planes as a function of the rigidity measured by the magnetic spectrometer. 
The helium, proton, electron and pion distributions are well separated thanks to the excellent MIP resolution. The upper and the lower 
bands (bounded black lines) represent the selections for the helium nuclei and protons respectively. Right panel: a calorimetric variable as a function of the particles 
rigidity. The black lines represent a constant efficiency selection defined in order to separate the hadronic and leptonic signals. 
} 
\label{fig3}
\end{figure}

A set of criteria based on the information provided by the sub-detectors described in the previous section 
was developed in order to select a clean sample of protons, helium and electrons from the data collected by the 
PAMELA instrument. 
Only events with a single reconstructed track were selected. The track was required to be located 
inside a fiducial volume bounded $0.15$ cm from 
the magnet cavity walls in order to avoid interaction with the magnetic
walls which could degrade the tracker performance. 
Protons and helium nuclei were selected by means of the 
ionization energy losses in the tracker and the ToF planes. Figure \ref{fig3} (left panel) shows 
the average ionization energy loss in terms of minimum ionizing particle (MIP)\footnote{Energy loss is expressed in 
terms of MIP that is the energy released by a particle which
mean energy loss rate in matter is minimum.} inside the silicon tracker planes.
Data were collected by the PAMELA instrument between 
July and December $2006$. The black lines represent a constant efficiency
selections on the proton (lower bands) and helium (upper bands) nuclei. 
No isotopic separation (proton/deuterium or $^3$He/$^4$He) was performed in this analysis. The dE/dx
selections provide a clean sample of helium nuclei and a sample of protons with 
a negligible positron contamination of the order of $10^{-4}$-$10^{-5}$ over the whole energy 
range. For more detail about the proton and the helium selection see \citep{protHE}.

Electrons were selected exploiting the PAMELA electromagnetic calorimeter. 
The main background is represented by galactic antiprotons (a few percent of the signal) and 
negative pions which are produced by the interaction of primary cosmic rays nuclei 
with the aluminum container that encloses the PAMELA instrument (few percent below $\sim$ 5 GV). 
Several selections based on the topological development of the particle 
shower were defined. Figure \ref{fig3} (right panel) shows the rigidity distribution of one 
calorimetric variable which was defined in order to emphasize the multiplication and the 
collimation of the electromagnetic shower. This variable represents the sum over 
all the calorimeter planes of the number of 
strip hit around a few centimeters from the shower axis.
Since the leptonic shower is more collimated 
than the hadronic one, electrons are characterized by 
higher values of this variable as shown in Figure \ref{fig3}. The black lines 
represent a constant efficiency selection defined in order to reject antiproton 
and negative pion contamination. 
A set of six calorimetric selections allowed an almost complete rejection of the 
antiproton and pion contamination 
in the rigidity range considered. The residual contamination was estimated using 
both simulated and flight data and was found to be less than one percent 
over the whole energy range. For more details about the electron selections and 
the estimation of the residual contamination see \citep{eletime}.

In order to reject reentrant albedo particles, the events were selected by imposing that the lower edge of the rigidity bin to
which the event belongs exceeds the critical rigidity, defined as $1.3$ times the geomagnetic cutoff rigidity  
computed in the St\"{o}rmer vertical approximation \citep{cutoff}

   \begin{figure}[t]
 \centering
 \includegraphics[width=19.5cm]{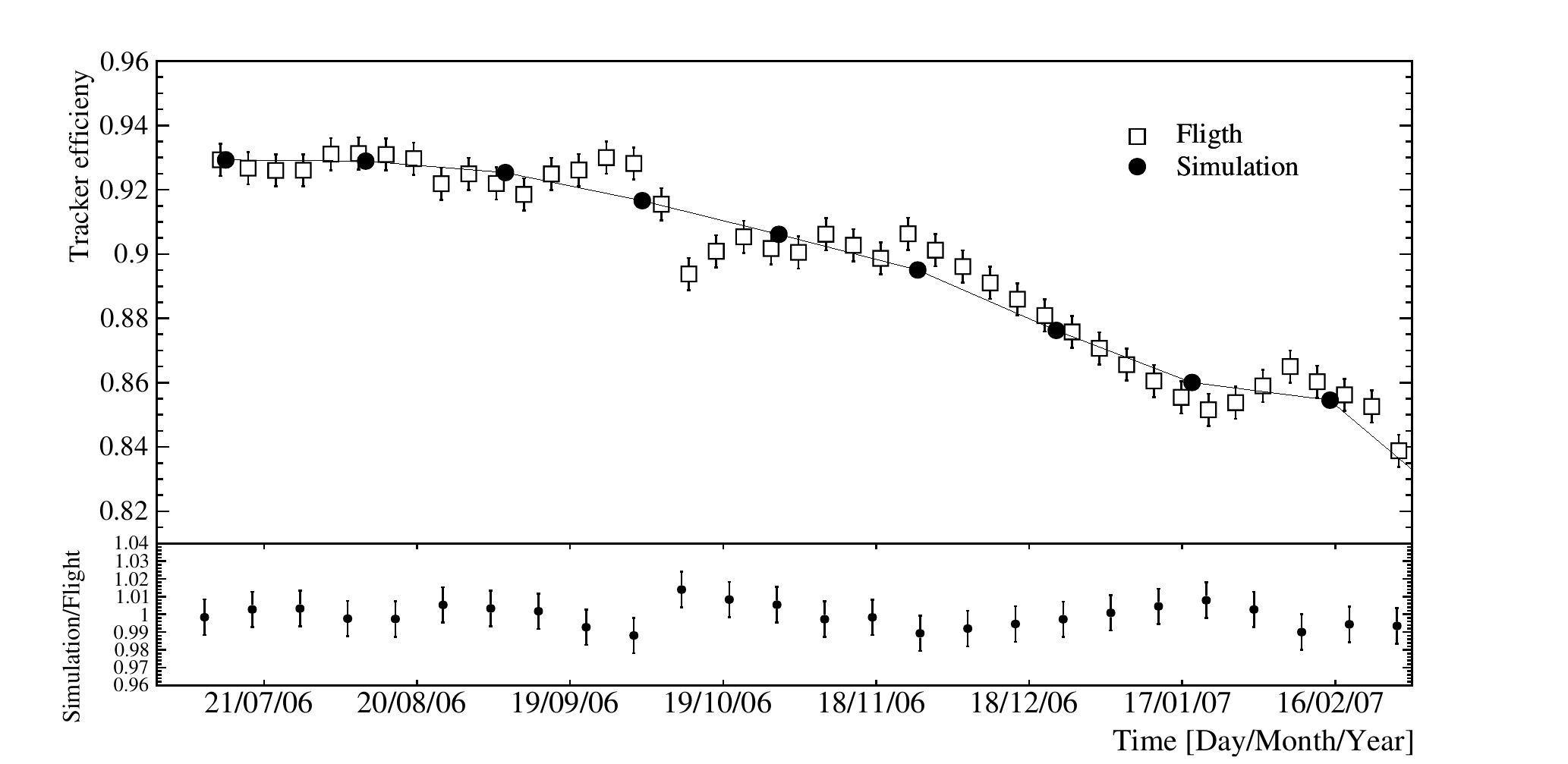}
 \caption{Top panel: the efficiency of the tracker selection as a function of time. The open squares  
 represent the weekly efficiency evaluated with flight data while the full circles represents the simulated efficiency averaged 
 on one month. The solid line connecting the simulated efficiencies represents the final values used for the flux calculation.
 Bottom panel: the ratio between the simulated and flight efficiencies. The ratio was calculated interpolating 
 the simulated efficiencies for the temporal division of the flight efficiency. }
 \label{fig4}
\end{figure}

The proton, helium and electron fluxes were finally calculated by dividing the number of particles by the selection 
efficiencies, the live-time and the geometrical factor. 
To avoid any biases which could introduce systematic temporal variation in the fluxes, the temporal evolution of the 
selection efficiencies was studied. The dE/dx selection was
found to have constant efficiencies during the whole time interval under analysis. Also the 
 calorimeter selection efficiencies were found constant over time. On the contrary the tracker selection 
 efficiency was found to decrease over time. This effect is ascribed to the random failure of few read-out chips of the 
 silicon mictrostrip detectors. The tracker efficiency was evaluated with two independent procedures over the period of time 
 from the beginning of data taking until May 2007:  
 \begin{itemize}
  \item  The PAMELA simulation 
 software (based on GEANT4 \citep{geant}) was used to generate an isotropic set of protons in the energy range under analysis. 
 The events reconstructed inside the instrumental acceptance
were used to measure the energy dependence of the tracker efficiency.
 The simulation toolkit reproduced  
 the flight configuration of the tracker planes and its temporal evolution. 
 Because of the huge computational time required to process all the different tracker configurations, 
 the simulated efficiency was evaluated with a temporal resolution of one month.  
 \item Non-interactive protons that do not produce a hadronic shower in the calorimenter 
 and release energy only along their track were selected with the calorimeter from flight data.
 These events were used to measure the tracker selection efficiency. 
 This procedure allows the efficiency to be estimated only within an integrated
 energy range with a lower threshold of few GeV.  
  The high statistics allowed to estimate the weekly integrated 
 efficiency during the period of time under analysis. 
 \end{itemize}
 The results  are displayed in Figure \ref{fig4} (top panel) where both
 the simulated (full circles) and the flight (open squares) tracker efficiency as a function of time are shown.  
 As can be noticed from the bottom panel of Figure \ref{fig4} an agreement of the order of $2-3 \%$ was found 
 between the simulated and the flight efficiencies over the whole 
 time interval. Because of the good agreement between the simulated and the flight efficiencies, in order 
 to minimize the statistical fluctuation, the final fluxes were calculated using the interpolated values of
 the simulated efficiencies, i.e. the solid line connecting the simulated efficiencies in Figure \ref{fig4}. The differences between the simulated and the 
 flight efficiencies were considered as a systematical uncertainties associated to the flux evaluation.

 The geometrical acceptance, i.e. the requirement of triggering and containment, at least
$1.5$ mm away from the magnet walls and the TOF-scintillator edge, was evaluated by simulating 
an isotropic flux of particles over the PAMELA detector. 
 A constant value of $19.9$ cm$^2$ sr was found above $\sim 1 $ GV decreasing at low energy due to 
 the increasing particle bending. The live time was provided by an on-board
 clock that timed the periods during which
the apparatus was waiting for a trigger. Because of the relatively short time spent by the
satellite at high geomagnetic latitudes, the total live time, and thus the collected statistics, was reduced
to about $10\%$ of the total value at $500$ MV.

 In order to study the temporal variation of the GCR flux during December $2006$  
 the particle intensities were normalized to the averaged flux measured during the calendar month before the event, i.e.  
 November $2006$. A constant linear fit was performed to the proton, helium and electron fluxes 
 between the $2006$ November 1 and 30. Then the fluxes 
 were normalized to these values. It was assumed that  
 for the duration of the Forbush event, changes due to the long-term solar modulation had a negligible effect on the GCR intensity.

  The Forbush decrease amplitude and recovery time were studied with  
  protons in nine rigidity intervals  between $0.4$ and $20$ GV. 
  The statistics allowed the proton flux to be measured with a time resolution of $3$ or $6$ hours up 
  to $5$ GV and with a time resolution of one day above $5$ GV. Because of the 
  limited statistics with respect to protons, the helium and electron fluxes 
  were evaluated with a two days time resolution.

\section{Results}
\label{sec:res}

\subsection{Intensity time profile}
\label{timeprofile}

\begin{figure}[t]
 \centering
   \includegraphics[width=18.5cm]{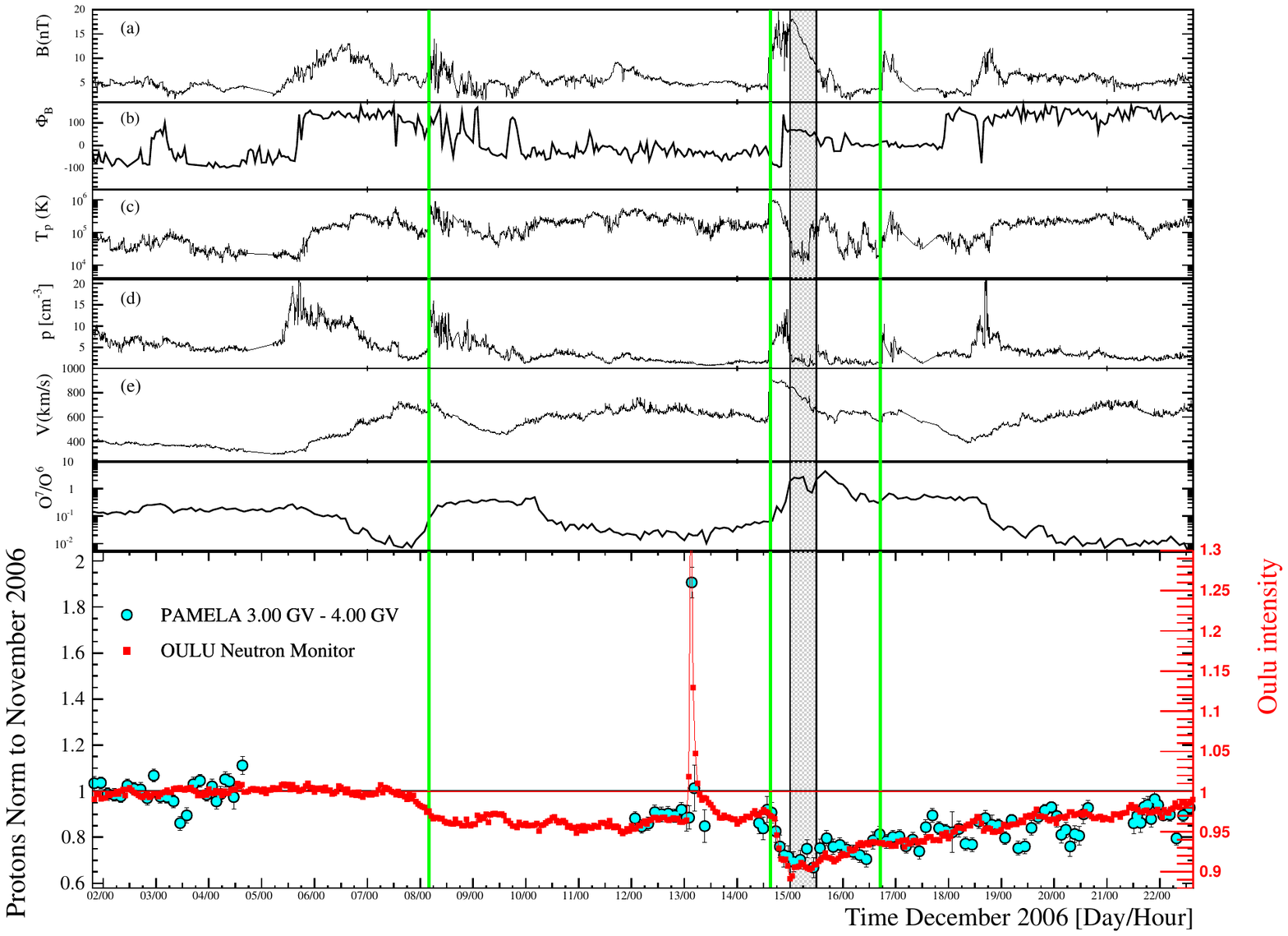}
 \caption{
  From the top: the HMF intensity (panel a), the HMF
azimuthal angles in GSE coordinates (panel b), the solar wind proton temperature (panel c), speed  (panel d) and density (panel e)
for $2006$ December $2$-$22$. Data are all $5$ minutes averages from the OMNI database (\url{https://omniweb.gsfc.nasa.gov/}). 
Panel (f) shows the solar wind ion charge state observation from the SWICS instrument on ACE, specifically, the O$^7$/O$^6$ ratio.
 Bottom panel: the three hours time resolution 
 proton intensity measured between $3$ and $4$ GV with the PAMELA instrument (full circles) 
 compared with the Oulu neutron monitor intensity averaged over one hour (full squares).  The solid horizontal line represents the 
 reference intensity on which data were normalized (November $2006$). Solid lines connecting the neutron monitor intensity are 
 displayed only to guide the eye. Vertical lines indicate times of shock passage while
 the grey shaded region indicates the passage of a magnetic cloud.    }

 \label{fig5}
\end{figure}

Figure \ref{fig5} places the proton intensity-time profile  measured by the PAMELA 
instrument and the neutron intensity measured by the Oulu neutron monitor in the context of near-Earth solar wind observations.
In detail, the bottom panel of Figure \ref{fig5} shows the proton intensity between $3$ and $4$ GV 
(full circles) measured by the PAMELA spectrometer with three hours temporal resolution compared with 
the hourly averaged neutron intensity measured by the Oulu neutron monitor (full squares). 
The PAMELA and the Oulu data are normalized to the average
November $2006$ intensity.
The other panels of Figure \ref{fig5} show from the top: the magnetic field intensity (panel a), the 
azimuthal angle in GSE coordinates (panel b), the solar wind proton temperature (panel c), speed  (panel d) and density (panel e).  
Panel (f) shows the solar wind ion charge state observation from the SWICS instrument on ACE, specifically, the O$^7$/O$^6$ ratio.

 As already pointed out in Section \ref{sec:intro}, neutron monitors respond to the GCR intensity variation 
over an integrated rigidity range with a lower threshold defined by their position on the Earth's surface. The lower threshold for the 
 Oulu neutron monitor is about $0.8$ GV. Moreover, neutron monitors are fixed on the Earth's surface while an instrument 
like PAMELA continuously orbits around the Earth. For these reasons the finer scale Oulu and PAMELA intensity variations 
during the Forbush decrease cannot
be directly compared. Nevertheless, Figure \ref{fig5} represents a useful crosscheck of the time evolution of the event. 
Moreover, the Oulu neutron monitor intensity fills in the gaps  when  PAMELA data are missing.

From Figure \ref{fig5} it can be noticed that the solar events of the $2006$ December $5$ and $6$ already produced 
a decrease in the GCR intensity that started at about the $1200$ UT of the $2006$ December $7$. 
The GCR decrease on December $7$ commenced just ahead of the arrival of a weak shock indicated by the first vertical line 
(the shock identification is from the shock database maintained by the University of Helsinki \url{http://ipshocks.fi/}).
 Since the GCR decrease clearly starts ahead of shock arrival, it is possible that the initial decrease could be  
  associated 
 with the corotating high speed stream through which the shock was propagating (see e.g. \cite{cane1,thomas}). 
 The extended GCR decrease without a recovery may then be associated with the persistent high speed flows that continue beyond 
 the time of the ground level event on $2006$ December $13$ in the Oulu data.
This suppression in the GCR intensity is also evident when the PAMELA data resume on $2006$ December $12$, with a decrease of about $10\%$
with respect to the average November GCR intensity, and continues to be present after the temporary increase associated with the $2006$ December 
$13$ solar event. 

The abrupt GCR decrease on December 14 commenced immediately following passage of the shock related to the solar event on $2006$ December 13. 
The shock (second vertical line) produced a geomagnetic storm sudden commencement at 1414 UT when it reached Earth 
(times from the CfA shock database, \url{https://www.cfa.harvard.edu/shocks/}). 
Signatures typical of ICMEs are evident for around four days 
following the shock, including intervals of depressed solar wind proton temperatures (panel c) and enhanced solar wind ion charge states (panel f). 
Following the shock on December 14, the GCR intensity declined and reached a minimum 
in the ICME with a magnetic cloud structure \citep{burg} present on $2006$ 
December $15$ (gray shaded band).
This shock and the ICME-associated structures immediately following are discussed in more detail by e.g. \cite{CME,CME3}.
At least one other shock passed by during this period of ICME-associated structures (third vertical line), observed by Wind (at 1734 UT) and 
ACE (1721 UT) on December 16.  The complexity of these structures indicates that this region is formed by the interaction of multiple ICMEs
but further analysis of these structures is beyond the scope of this paper. 

 The observations suggest that the decrease commenced on the $2006$ December $13$ then added to the larger decrease commencing on December 14, though 
the contributions of the two decreases cannot be disentangled.  Therefore, in this study the decrease commencing on December 14 is treated here 
as being entirely due to the passage of the shock and ICME on December $14$-$15$.  

\begin{figure}[t]
 \centering
 \includegraphics[width=19.5cm]{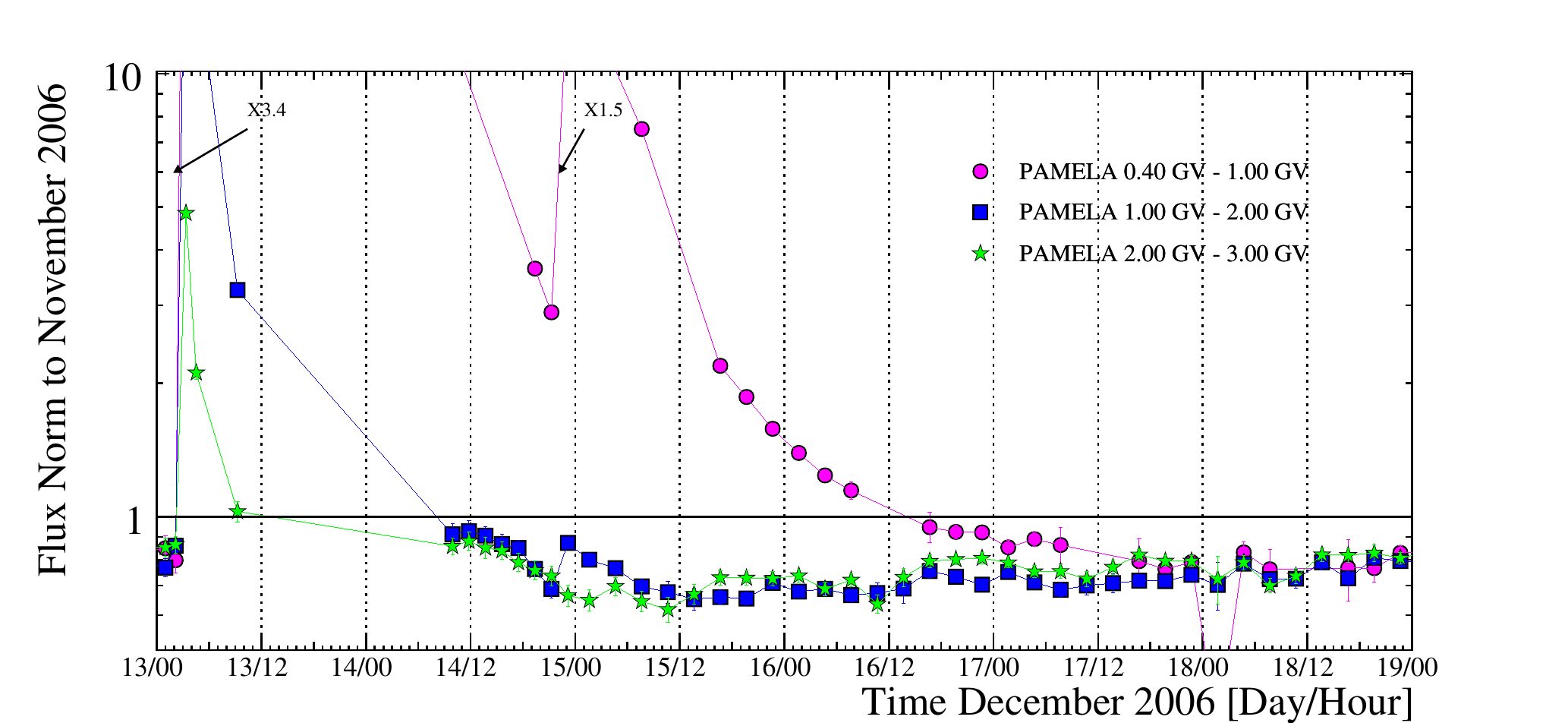}
 \caption{The three hour time resolution proton intensity measured with the PAMELA instrument. 
 Three different rigidity intervals are displayed
 between the $2006$ December $13$ and $18$. The solid horizontal line represents the 
 reference intensity to which the data were normalized, i.e. the average GCR intensity measured during November $2006$ by PAMELA. Solid 
 lines connecting each point are displayed only to guide the eye. The arrows indicate the time of the flares at the Sun.}
 \label{fig6}
\end{figure}

Figure \ref{fig6} shows the proton intensity-time profile measured by the PAMELA instrument
(three hour time resolution) over three different rigidity intervals. The intensity profile measured at the lowest rigidity ($0.4$-$1$ GV)  
shows the arrival of the solar energetic particles around $0300$ UT on the $2006$ December $13$ 
and again around $2200$ UT on the  $2006$ December 14 associated with the flares and fast CMEs at these times described above.
Missing data are due to on-board system reset caused by the 
high trigger rate that occurred  during the solar events.
Solar energetic particles between $0.4$ and $2$ GV were visible following both solar events, while between $2$ - $3$ GV, only 
the $2006$ December 13 solar event produced a visible increase. 
Above $1$ GV the Forbush decrease started around $1200$ to $1500$ UT on the $2006$ December $14$, 
associated with the arrival of the interplanetary shock and reached the 
minimum intensity during the first half of $2006$ December $15$ at $2$-$3$ GV, $1$-$2$ GV protons being dominated
by solar particles at this time. 
Below 1 GV, solar energetic particles continue to dominate, and the intensity only falls below the pre-event background,
indicating a decrease in the GCR intensity, approximatively three days later at
around $2100$ UT on the $2006$ December $16$ and 
reached its minimum intensity around $0000$ UT on the $2006$ December $18$.


\begin{figure}[t]
 \centering
 \includegraphics[width=19.5cm]{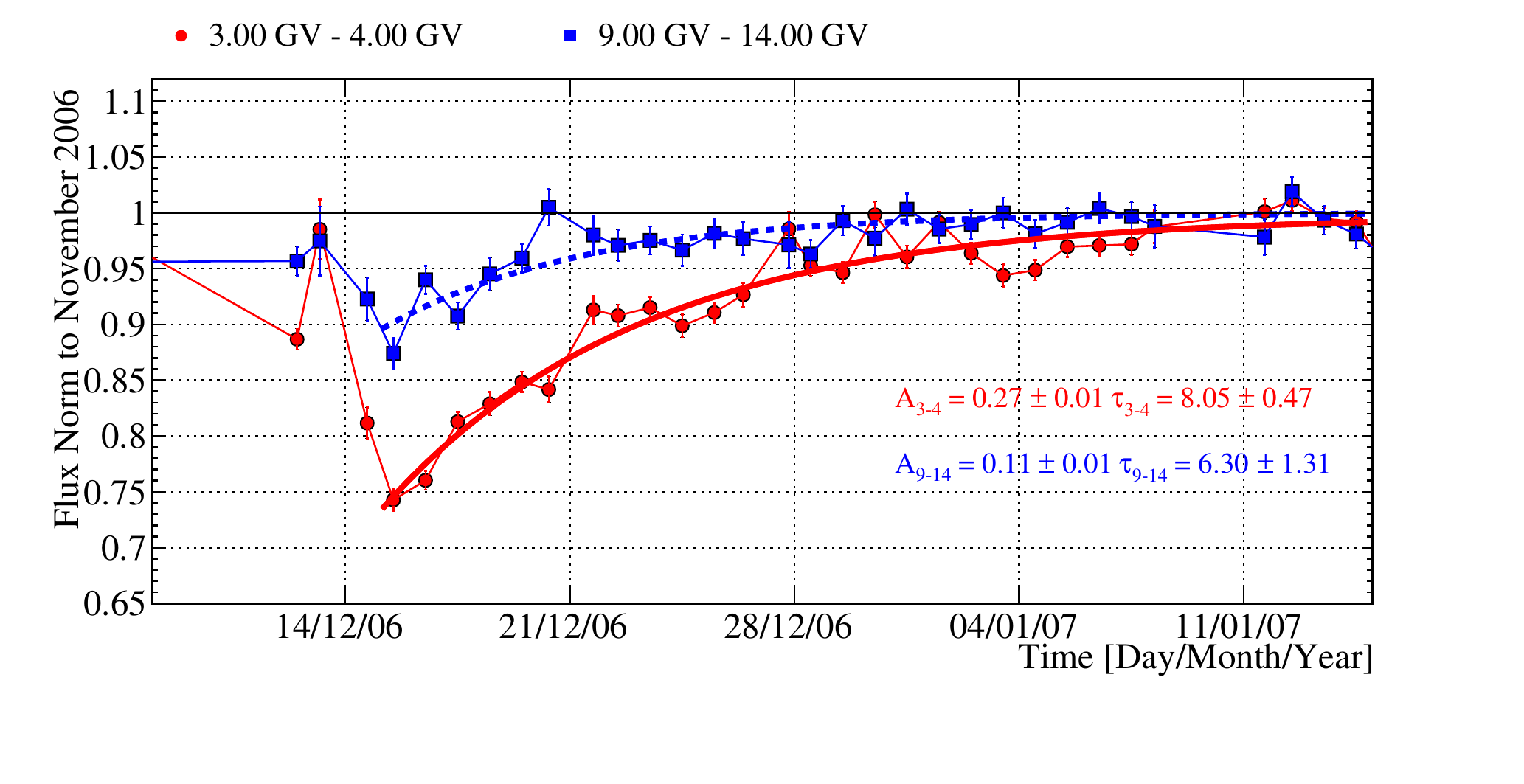}
 \caption{The proton intensity over time measured by the PAMELA instrument between $2006$ December $14$  and 
 $2007$ January $16$. Each point represents one day of data taking. Two rigidity intervals are showed: 3-4 GV (circle points) 
 and 9-14 GV (square points). The solid and the dotted lines represent a fit performed with Equation \ref{fit} to the 3-4 GV and 9-14 GV interval respectively. 
 The values for the amplitude $A$ and the recovery time $\tau$ obtained from the fit are also showed. }
 \label{fig7}
\end{figure}

\subsection{Amplitude and recovery time rigidity dependence}
\label{ampl}

The amplitude and the recovery time of the Forbush decrease commencing on $2006$ December $14$ was studied by fitting the 
time profile of the daily average proton fluxes, $I(t)$, in nine different rigidity intervals. The following function was used (e.g. \cite{us1,neutron}): 
\begin{equation}
\label{fit}
 I(t) = 1-A \ e^{-\frac{t-t_{0}}{\tau}}
\end{equation}

\begin{figure}[t]
 \centering
  \includegraphics[width=19cm]{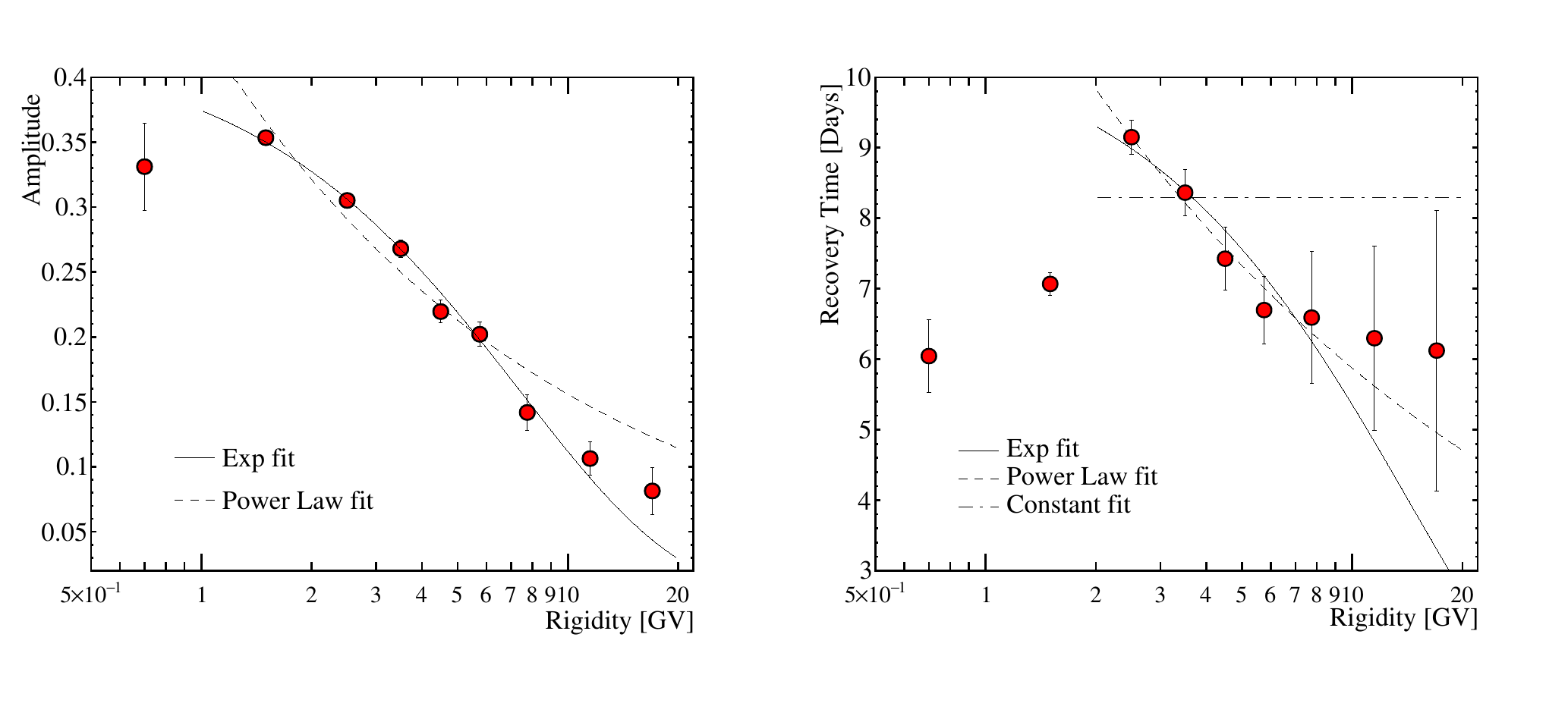}

 \caption{Left panel: the rigidity dependence of the Forbush
 decrease commencing the $2006$ December $14$ as measured by the PAMELA instrument. Each value 
  results from an exponential fit (Equation  \ref{fit}) performed on the daily average time profile of the 
  proton intensity. A total of nine rigidity intervals between $0.4$ and $20$ GV were studied. The solid and dashed 
  black line represent 
  an exponential (Equation \ref{eq2})  and a power law fit (Equation \ref{eq3}) respectively performed on the amplitude as a function of the rigidity. 
  The parameters and the associated errors of the parameters obtained from the fit are displayed in Table \ref{tab:fit}. 
  Right panel: as the left panel for the recovery time $\tau$.   
  The dotted-dashed line represents a fit with a constant value (Equation \ref{eq4}).  
}
 \label{fig8}
\end{figure}

The free parameters are the amplitude $A$ of the decrease with respect to the reference flux
and the recovery time $\tau$. The absolute
reference time $t_{0}$, which represent the starting time of the Forbush decrease, is fixed. Both $\tau$ and $t_0$ are 
expressed in days.
As already discussed in Section \ref{timeprofile}, because of the prolonged 
presence of the solar particles, between $0.4$ and $1$ GV the Forbush decrease (measured by the PAMELA instrument)
starts with three days of delay with respect to the higher rigidities.
For this reason above $1$ GV the fit was performed between the $2006$ December $15$ at $0000$ UT and the $2007$ January $31$ at $2400$ UT
while below $1$ GV the fit was performed starting from the $2006$ December $18$ at $0000$ UT. However, by assuming that 
the Forbush decrease started at the same time for all the rigidities, the amplitude and the 
recovery time  below 1 GV were also calculated with $t_0$ set to $0000$ UT on $2006$ December $15$.
Figure \ref{fig7} shows the result of these fits to the daily average proton flux 
 normalized to the November 2006 proton intensities for two different rigidity intervals. The full circles represent protons measured 
 between 3-4 GV while the full squares represent protons measured between 9-14 GV. 
 The solid and the dotted lines are fits performed with Equation \ref{fit} to the 3-4 GV and 9-14 GV intervals respectively. 
 In order to study the rigidity dependence of the amplitude and the recovery time a total of nine rigidity intervals were studied 
 between $0.4$ - $20$ GV.
Figure \ref{fig8} (left panel) shows the rigidity dependence of the Forbush decrease amplitude obtained with the fitting procedure
while the right panel displays the rigidity dependence of the recovery time $\tau$.
A general decreasing trend with increasing rigidity is observed 
 both for the amplitude and the recovery time. 
However, it can be noticed that the first point of the amplitude distribution and the first two points for the recovery time distribution 
are in disagreement with the decreasing trend. This could point to a real physical effect or could be a limitation of the fitting procedure 
due to the contamination of the solar energetic particles that biases the fit results.

The rigidity dependence of the amplitude and the recovery time 
were fitted by means of an exponential and a power law:
\begin{equation}
\label{eq2}
 a \ e^{- \alpha R}
\end{equation}
\begin{equation}
\label{eq3}
 b \ R^{- \beta}
\end{equation}
 where $R$ is the rigidity and $a$, $\alpha$, $b$ and $\beta$  are free parameters to be determined. In addition the rigidity dependence of the 
 recovery time was also fitted with a constant: 
 \begin{equation}
\label{eq4}
 \eta
\end{equation}
The solid black lines on  Figure \ref{fig8} (left and right panels) 
represent the exponential fits while the dotted lines refer to the power law fits. The dashed-dotted line in Figure \ref{fig8} (right panel)
represent the fit with a constant. The first point of the amplitude 
distribution and the first two points of the recovery time distribution were not used in the fits. The results of the 
fits are displayed in Table \ref{tab:fit}.
From the $\chi^2$/NDF (number of degree of freedom) it can be noticed that the rigidity dependence of the Forbush decrease amplitude is better described with 
an exponential while the recovery time is well fitted with both a power law and an exponential fit. 
On the other hand the hypothesis of rigidity independence of the recovery time is heavely disfavoured from the $\chi^2$/NDF (see Table \ref{tab:fit}). 
The rigidity dependence of the recovery observed by PAMELA for the December $2006$ Forbush events was 
previously observed by \cite{neutron} combining observations from several neutron monitor stations (median rigidity between $10$ and 
$30$ GV) and the MUG muon telescope in Finland (mean rigidity of $55$ GV) and obtaining $\alpha = 0.023 \pm 0.007$.

It is generally thought that the main drivers of the recovery time are the decay of the interplanetary disturbance 
and to a lesser extent on the transport parameters of GCRs which would imply that the recovery time is not energy 
dependent (e.g. \cite{Loockwood,Wibberenz}).  However, \cite{mulder} argued that the Forbush decrease recovery time is affected by the interplanetary magnetic field polarity and thus
the drift of GCRs in the heliosphere, which implicitly depends on energy. 
Our results for the December $2006$ Forbush event sustain this energy dependence of the recovery. 
Other recent results (e.g. \cite{neutron,Zhao}) found both events with and without energy dependence of the recovery
concluding that this dependence is strongly related to the features of the solar disturbance causing the Forbush decrease.

\subsection{Proton-electron-helium comparison}

\begin{figure}[t]
 \centering
 \includegraphics[scale=0.42]{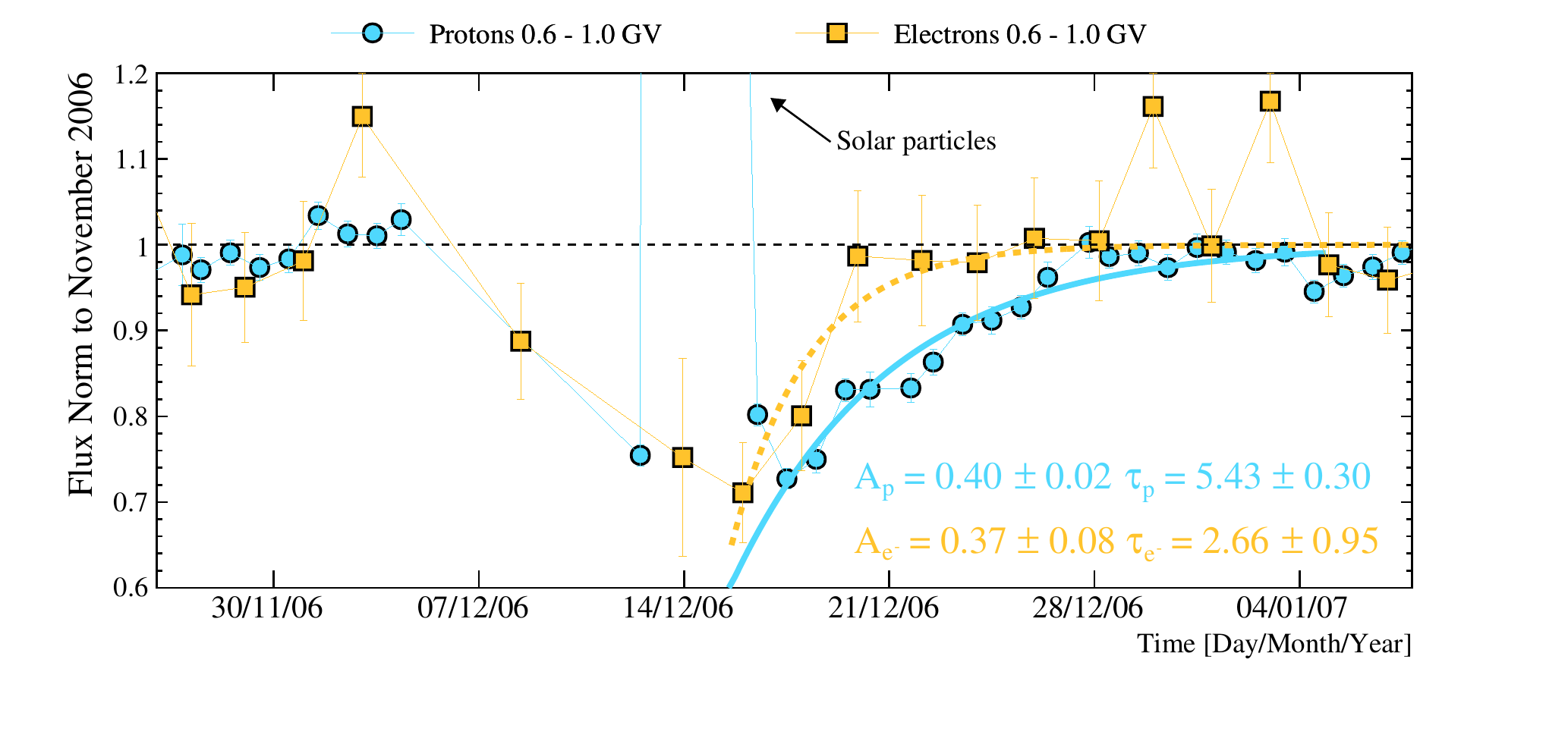}  \includegraphics[scale=0.42]{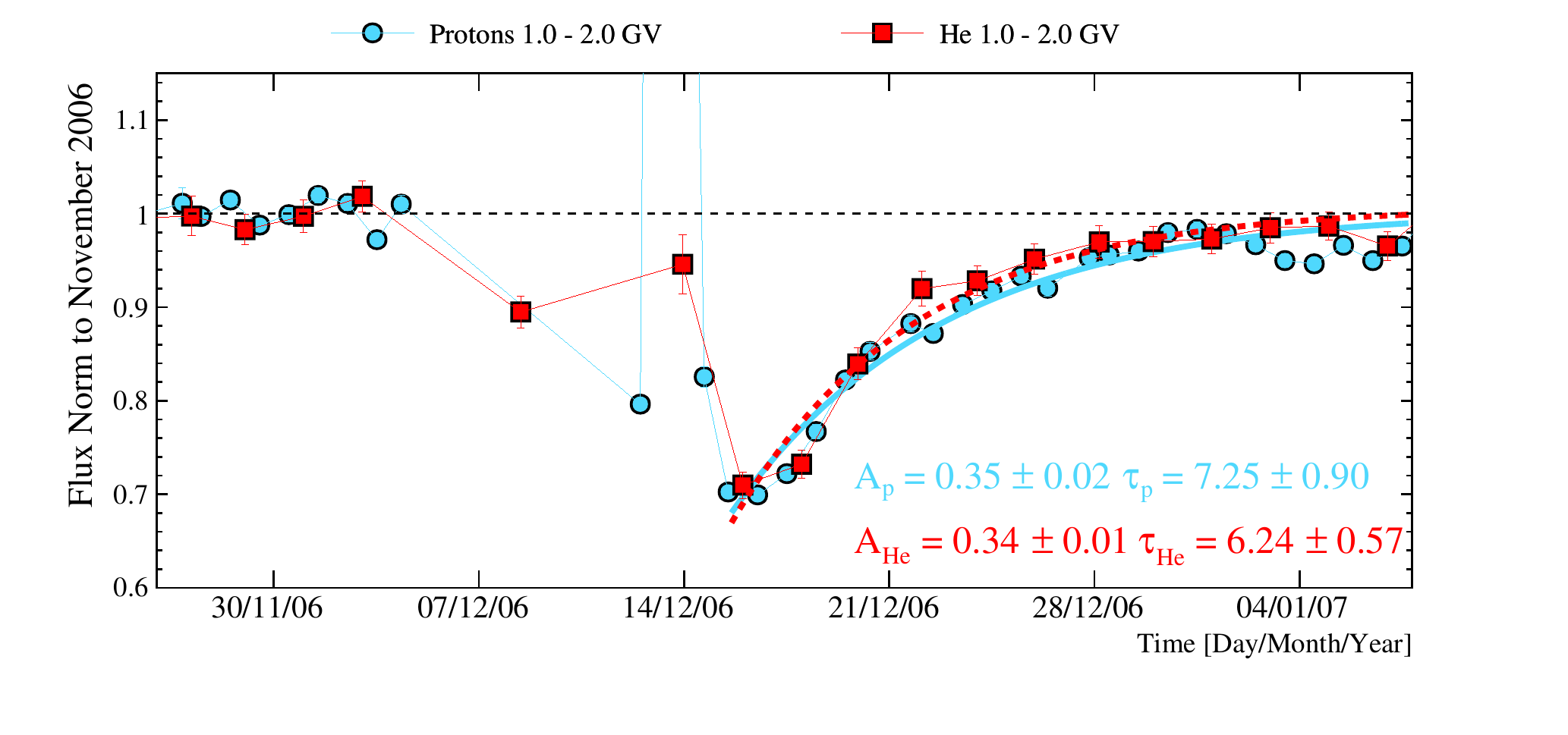} \\
 \includegraphics[scale=0.42]{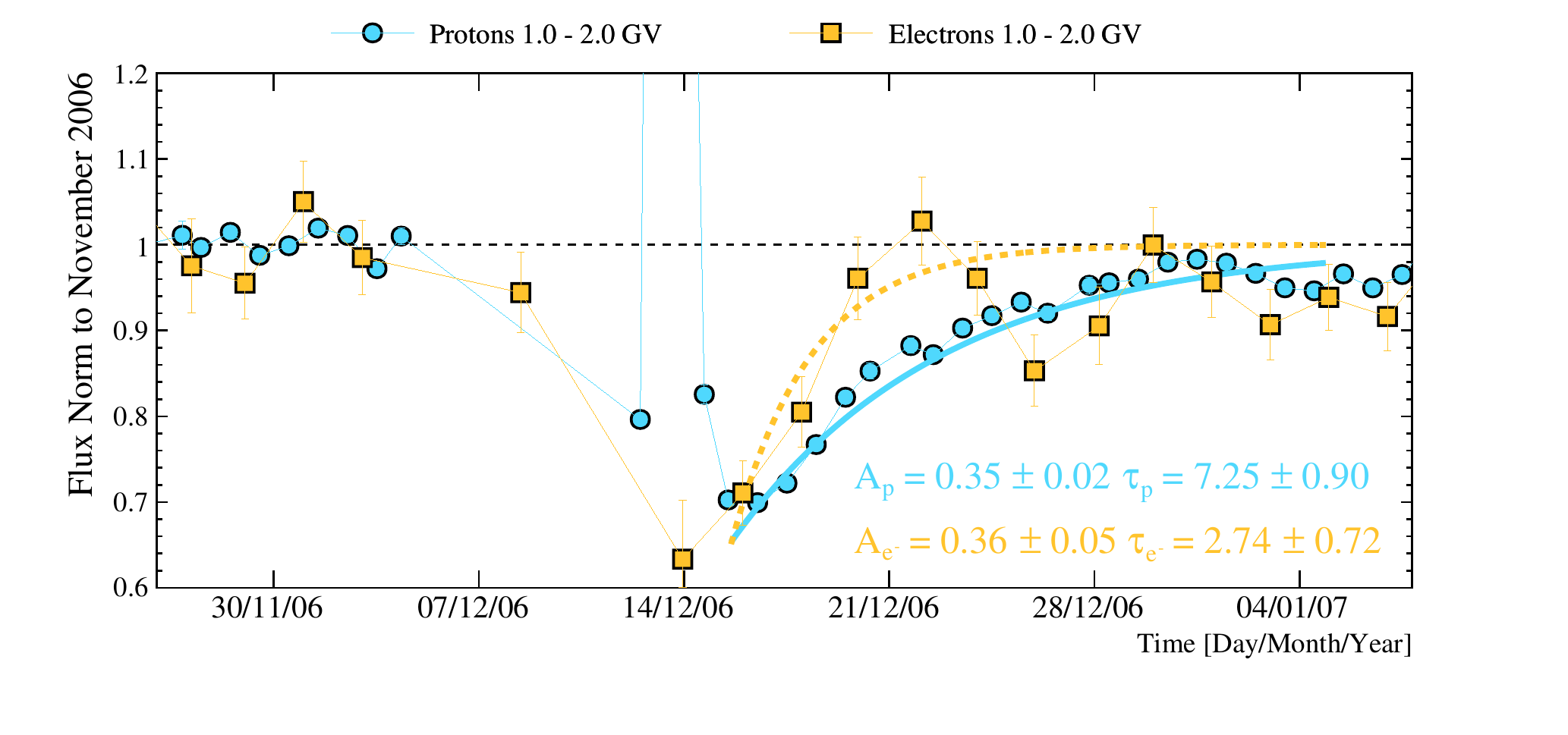} \includegraphics[scale=0.42]{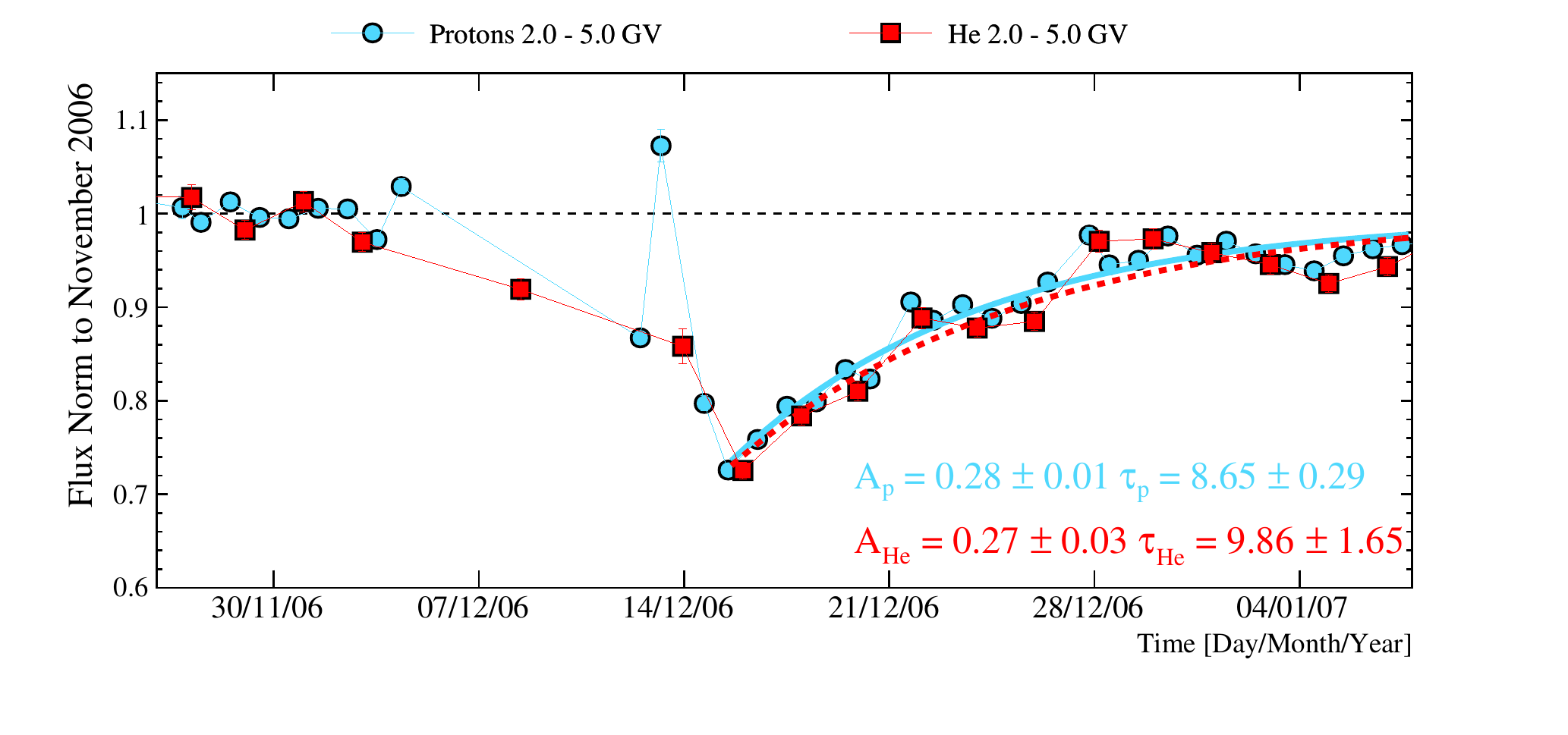} \\
 \includegraphics[scale=0.42]{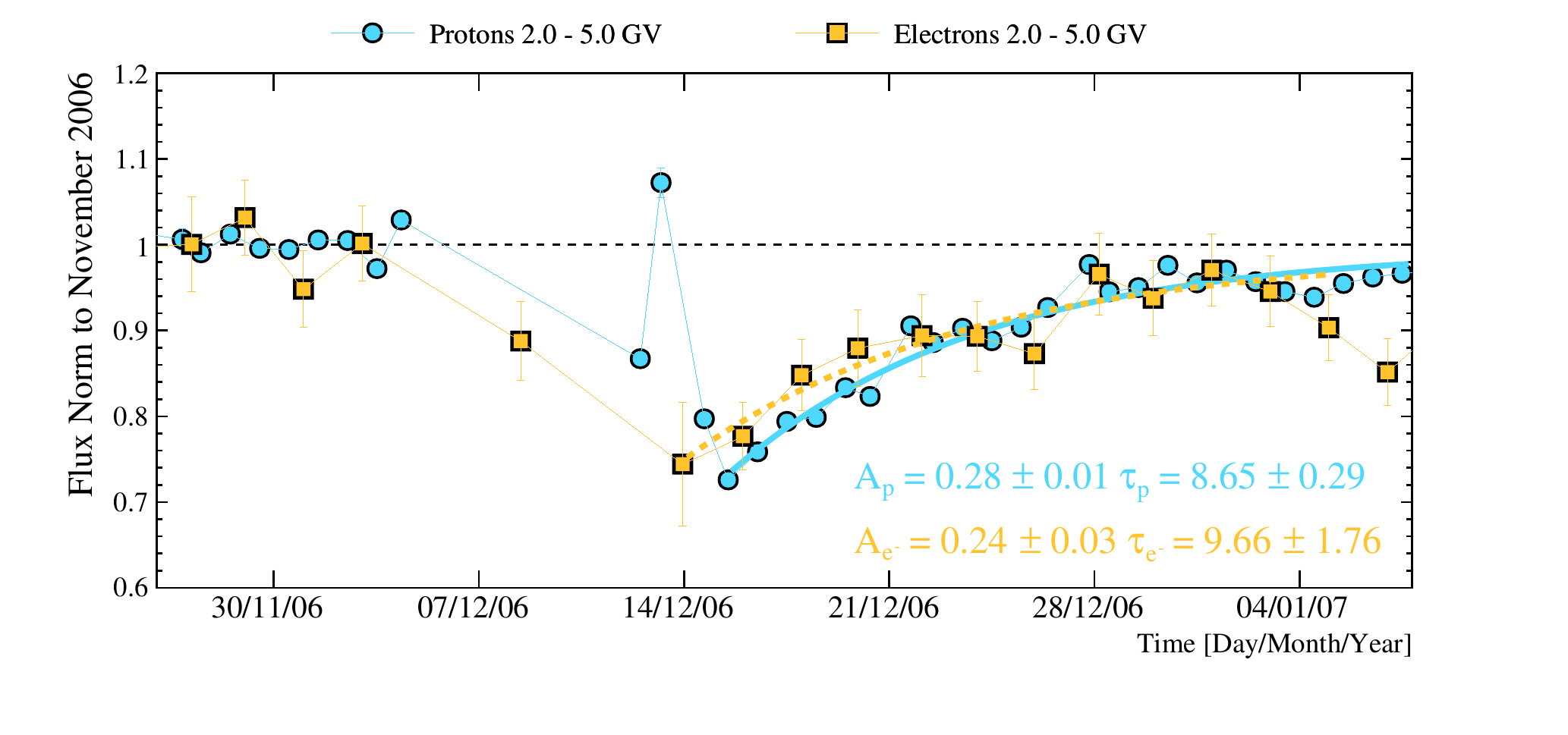} \includegraphics[scale=0.42]{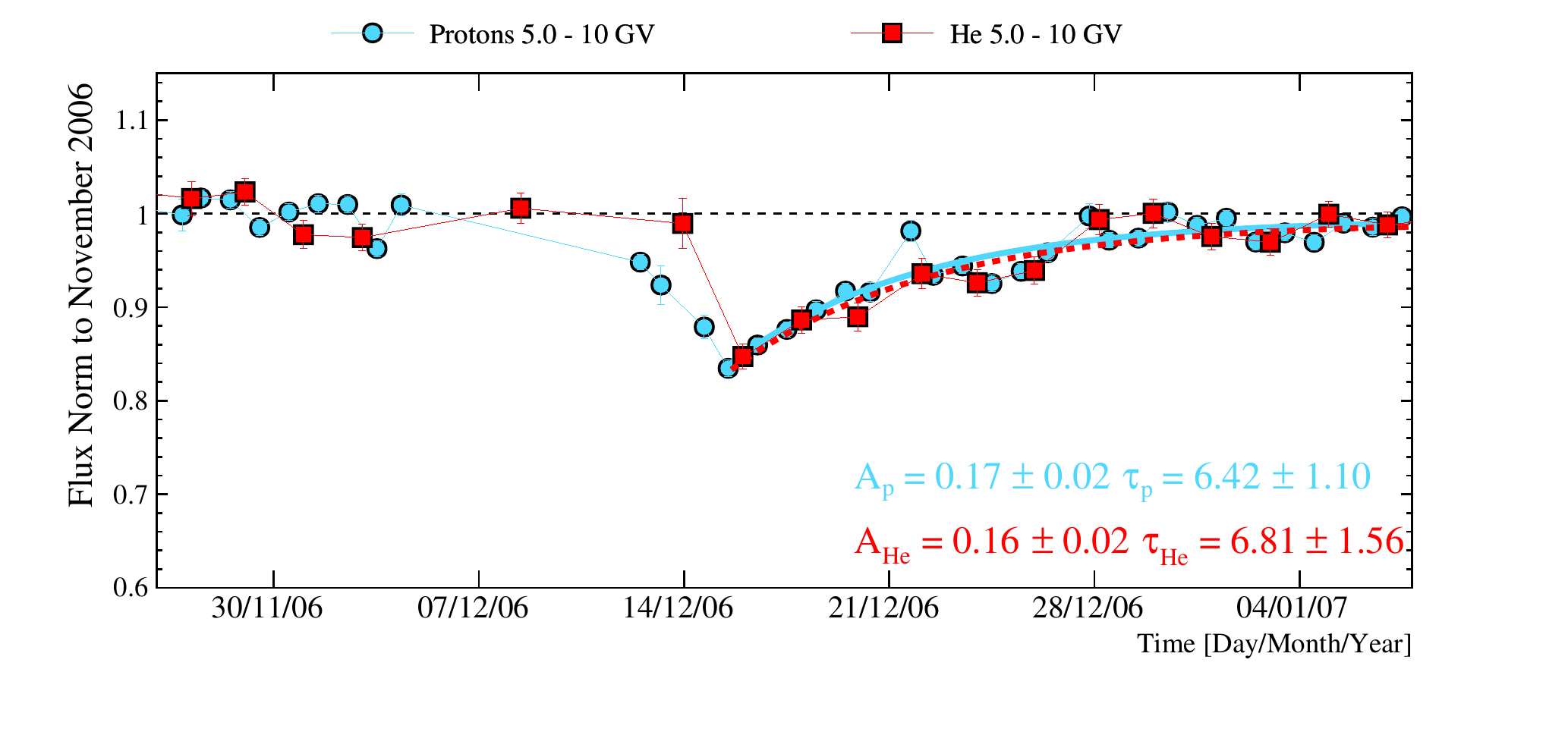}
 \caption{Left panels: proton (full circles) and electron (full squares) time profile intensities for three different rigidity ranges during the 
 December $2006$ Forbush decrease. 
 The proton intensity is averaged over one day while the electron is averaged over two days. Right panels: as for the left panel with 
 proton (full circles) and helium 
 (full squares).  The solid horizontal lines represent the 
 reference intensity on which data were normalized (November $2006$). The solid and the dotted lines represent exponential fits performed with Equation \ref{fit} respectively to 
 proton and electron in the left panels and to proton and helium in the right panels. In each panels 
 the amplitude $A$ and the recovery time $\tau$ obtained with the exponential fit are reported. In both panels the lines connecting each point 
 are displayed only to guide the eye.}
 \label{fig9}
\end{figure}

As already discussed in Section \ref{sec:intro} the PAMELA instrument allows the Forbush decrease to be compared for 
different particle species. The proton intensity over time was compared to the electron and the helium intensities
in order to highlight possible differences in the amplitude and the recovery time.
The three left panels of Figure \ref{fig9} show the comparison between the daily averaged proton intensity (full circles) and the 
two days average electron fluxes (full squares). In order to increase the limited electron statistics with respect to the 
analysis described in Section \ref{ampl}, the lower limit of the first rigidity interval was increased from $0.4$ GV to
$0.6$ GV. Taking into account the discussion in Section \ref{sec:data} this was equivalent to increasing 
 the total live time spent by the satellite at geomagnetic latitude 
suitable for detecting galactic particles. An overall increase of the live time of about $40 \%$ was achieved.
Moreover, the third rigidity interval 
was extended up to $5$ GV in order to increase the statistics.

\begin{figure}[t]
 \centering
 \includegraphics[scale=0.82]{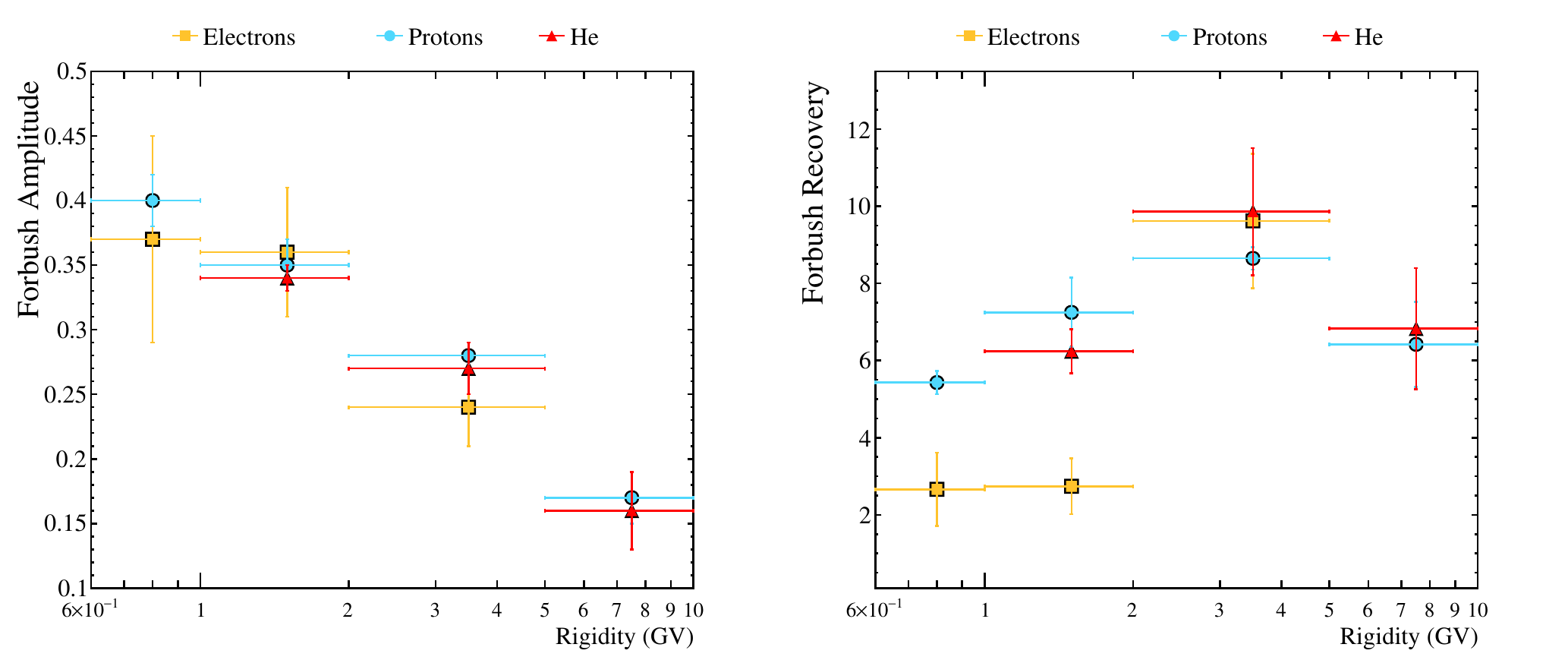}  
 \caption{ Left panel: the rigidity dependence of the December 2006 Forbush decrease amplitude for protons (circles), electrons (squares) and helium nuclei (triangles). 
 Right panel: same as left panel for the recovery time.}
 \label{fig10}
\end{figure}

The three right panels of Figure \ref{fig9} 
show the proton intensity (circle points) compared with the helium intensity (square points). 
Because of the energy losses inside 
the apparatus, the PAMELA instrumental limit 
for helium detection is about $0.8$ GV. For this reason the first 
rigidity interval was chosen to be $1$-$2$ GV while the last one was $5$ - $10$ GV.
In order to emphasize possible differences in the amplitude or the recovery time an exponential fit  
(Equation \ref{fit}) was performed on the proton, electron and helium intensity profiles over time. The solid lines and the 
dotted lines in the left panels represent the proton and electron fits respectively, while in the right panels they represent the 
proton and helium fits respectively. 
The amplitude and the recovery time resulting from the fits as a function of rigidity are shown on left and right panel of Figure \ref{fig10} respectively.

As can be seen in Figure \ref{fig10} the helium and the proton amplitude and recovery time are in agreement within the errors
for each rigidity interval. On the contrary electrons show a faster 
recovery time with respect to the protons for the first two rigidity intervals while having the same amplitude.
The recovery time shows a better agreement between 
protons and electrons in the last rigidity interval ($2$-$5$ GV).
This differences could be interpreted as an effect of the GCR propagation inside the heliosphere, in particular 
the charge sign dependence introduced by drift motions \citep{drift}. 
In particular, since the near-Earth ICME extent 
was estimated to occupy a significant solid angle in the heliosphere, i.e. $74^{\text{\textopenbullet}}$   
in latitude and $117^{\text{\textopenbullet}}$ in longitude \citep{CME} with respect to the heliospheric equator, possibly 
global drift motions may play a role in the recovery phase. 
In fact, during $A < 0$ epochs\footnote{In the 
 Sun magnetic field the dipole term nearly always dominates the magnetic field of the solar wind.
A is defined as the projection of this dipole on the solar rotation axis.} such 
as in the declining phase of cycle 23, when the heliospheric magnetic field is directed toward the Sun in the northern hemisphere,
negatively charged particles undergo drift motion mainly from the polar to the equatorial
regions while positively charged particles drift
mainly in the opposite directions. 



\begin{table}[t]
\centering
\caption{Fitted parameters for the rigidity dependence of $A$ and $\tau$ performed with Equations \ref{eq2} and \ref{eq3}. } \label{tab:fit}
\begin{tabular}{ccc}
\tablewidth{0pt}
\hline
\hline
		       & Amplitude & Recovery Time\\
\hline
$\chi^2_{\text{exp}}$/NDF  &  $9.2/6$     	& $5.6/5$     \\
$a$ 		           &  $0.43 \pm 0.01$   &  $10.7 \pm 0.5$     \\
$\alpha$ 	           &  $0.134 \pm 0.007$ &  $0.07 \pm 0.02$    \\
$\chi^2_{\text{power law}}$/NDF      &  $41/6$     	& $1.5/5$     \\
$b$ 		           &  $0.43 \pm 0.01$  	&   $12.2 \pm 0.9$   \\
$\beta$		           &  $0.45 \pm 0.02$   &   $0.31 \pm 0.06$ \\
$\chi^2_{\text{constant}}$/NDF      &  $ $     	& $34/6$     \\
$\eta$		           &  $ $   &   $8.2 \pm 0.2$ \\

\hline

\end{tabular}
\end{table}

Moreover, as discussed above, at least two shocks and ICME combined during the period of the December $2006$ 
Forbush decrease , which would presumably occupy a larger extent than the single shock/ICME.
Thus, considering the ICMEs topology just discussed, it can 
be argued that the equatorward electron drift direction would be expected to help fill in the Forbush decrease resulting in a 
faster recovery time with respect to the protons which 
drift mainly in the opposite direction and would experience a longer suppression. 
This result is in agreement with  \cite{LeRoux}  who predicted that the recovery time is longer for positively charged particles
when the polarity of the solar magnetic field is negative solving a 2D
transport equation, including adiabatic cooling and particle
drift.
The charge sign dependence introduced by drift motions in the global solar modulation of GCRs, i.e. the ratio of GCRs intensities during opposite 
polarity cycle, is expected to have a maximum effects at around between $300$-$500$ MV becoming 
less than few $\%$ at $5$ GV (see e.g.. \cite{posfractime,Felice, Nndanganeni}). This may explain why the differences in the recovery time between electrons and protons 
are greater at the lowest rigidities and tend to reduce as the rigidity increases.

\section{Conclusion}
For the first time a Forbush decrease ($2006$ December $14$) was extensively studied using observations of GCR 
in space with the PAMELA instrument. The proton observations have sufficient statistics to make it possible to study the temporal evolution 
of the event with three hours time resolution for different rigidity intervals between $0.4$ - $20$ GV.
The rigidity dependence of the amplitude and the recovery time were investigated over nine different 
intervals. The amplitude of the Forbush decrease was found to decrease as the rigidity increased. An exponential fit 
 describes well the rigidity dependence for the amplitude. The recovery time shows an increasing trend below one GV 
which could be either a limitation of the fitting procedure 
due to the contamination of the solar energetic particles or a real physical effect. Above 1 GV a general 
decreasing trend is found. Both the exponential and power law well fit this distribution.  

For the first time the PAMELA observation  allowed to study the behavior of different particle species during a Forbush decrease.
In particular protons, helium nuclei and electrons were compared. 
The proton and the helium nuclei amplitude as well as the recovery time were found in good agreement while electrons showed
on average a faster recovery time which tended to approach the proton recovery time as the rigidity increased.
This behavior could be interpreted as a charge-sign dependence due to the different 
global drift pattern between protons and electrons. 

 The results discussed in this paper will be available at the Cosmic Ray Data Base of the ASI Space Science Data Center 
  (\url{http://tools.asdc.asi.it/CosmicRays/chargedCosmicRays.jsp}).

We acknowledge partial financial support from The Italian Space Agency (ASI) under the program ``Programma PAMELA - attivit\'a scientifica di analisi dati in fase E".
We also acknowledge support from Deutsches Zentrum fur Luft- und Raumfahrt (DLR), The Swedish
National Space Board, The Swedish Research Council, The Russian Space Agency (Roscosmos) and Russian Ministry of Education and
and NASA Supporting Research Grant $13-SRHSPH13 20075$. M. S. Potgieter acknowledge the partial
financial support from the South African Research Foundation (NRF) under the SA-Italy
Bilateral Programme. I. G. Richardson acknowledges support from the ACE mission.



\end{document}